\documentclass[letterpaper, 10 pt, conference]{ieeeconf}  % Comment this line out
\IEEEoverridecommandlockouts                              % This command is only
\usepackage{amsmath} % assumes amsmath package installed
\usepackage{amssymb}  % assumes amsmath package installed
\usepackage{physics}
\usepackage[caption=false,subrefformat=parens]{subfig}
\usepackage[ruled]{algorithm2e}
\usepackage{booktabs}
\usepackage{threeparttable}
\usepackage{mathtools}
\DeclarePairedDelimiter\sbrac{(}{)}

\DeclarePairedDelimiter\bbrac{\{}{\}}
\newcommand{\sbr}[1]{\sbrac*{#1}}

\newcommand{\bbr}[1]{\bbrac*{#1}}

\def\calX{\mathcal{X}}
\def\calU{\mathcal{U}}

\def\st{\mathrm{ s.t. }}

\newtheorem{assumption}{Assumption}
\newtheorem{theorem}{Theorem}

\title{\LARGE \bf
On the Stability of Datatic Control Systems
}

\author{Yujie Yang, Zhilong Zheng, and Shengbo Eben Li$^*$% <-this % stops a space
\thanks{This study is supported by National Key R\&D Program of China with 2022YFB2502901, and NSF China under 52221005. It is also partially supported by Tsinghua University Initiative Scientific Research Program. Y. Yang and Z. Zheng contributed equally to this work. All correspondence should be sent to S. Li with email: lishbo@tsinghua.edu.cn}
\thanks{Yujie Yang, Zhilong Zheng, and Shengbo Eben Li are with School of Vehicle and Mobility and State Key Lab of Intelligent Green Vehicle and Mobility, Tsinghua University, Beijing, 100084, China {\tt\small \{yangyj21,zheng-zl22\}@mails.tsinghua.edu.cn,  lishbo@tsinghua.edu.cn}}}%

\begin{document}

\maketitle
% \thispagestyle{empty}
% \pagestyle{empty}

% for showing page number
\thispagestyle{plain}
\pagestyle{plain}

%%%%%%%%%%%%%%%%%%%%%%%%%%%%%%%%%%%%%%%%%%%%%%%%%%%%%%%%%%%%%%%%%%%%%%%%%%%%%%%%
\begin{abstract}
The development of feedback controllers is undergoing a paradigm shift from \textit{modelic} (model-driven) control to \textit{datatic} (data-driven) control. Stability, as a fundamental property in control, is less well studied in datatic control paradigm. The difficulty is that traditional stability criteria rely on explicit system models, which are not available in those systems with datatic description. Some pioneering works explore stability criteria for datatic systems with special forms such as linear systems, homogeneous systems, and polynomial systems. However, these systems imply too strong assumptions on the inherent connection among data points, which do not hold in general nonlinear systems.
This paper proposes a stability verification algorithm for general datatic control systems called $\eta$-testing. Our stability criterion only relies on a weak assumption of Lipschitz continuity so as to extend information from known data points to unmeasured regions. This information restricts the time derivative of any unknown state to the intersection of a set of closed balls. Inside the intersection, the worst-case time derivative of Lyapunov function is estimated by solving a quadratically constrained linear program (QCLP). By comparing the optimal values of QCLPs to zero in the whole state space, a sufficient condition of system stability can be checked.
We test our algorithm on three datatic control systems, including both linear and nonlinear ones. Results show that our algorithm successfully verifies the stability, instability, and critical stability of tested systems.
\end{abstract}

%%%%%%%%%%%%%%%%%%%%%%%%%%%%%%%%%%%%%%%%%%%%%%%%%%%%%%%%%%%%%%%%%%%%%%%%%%%%%%%%
\section{Introduction}
The historical evolution of control spans over two millennia, with early ingenious devices such as Ctesibius' feedback mechanism designed to regulate water clocks and recent industrial applications such as launch vehicle and spacecraft control in the Apollo lunar landing program~\cite{aastrom2014control}.
Since the first industrial revolution, the understructure of designing feedback controllers can be traced from classical control theory to modern control theory, both of which regard stability as a fundamental property of closed-loop systems. Generally speaking, stability refers to the ability of a control system to stay in the vicinity of an equilibrium point.
Classical control theory utilizes the Laplace transform to model systems as transfer functions. In classical control theory, typical methods to stabilize a system include pole placement, root locus method, and frequency response analysis. They have found extensive applications in mechanical devices, electrical systems, and chemical processes~\cite{tsien1954engineering, wiener2019cybernetics}.
Modern control theory represents a revolutionary shift from transfer functions in the frequency domain to state-space representations in the temporal domain, which performs system modeling, structure transformation, modal analysis, and controller synthesis using linear algebra techniques. This control theory is featured with modeling systems as state-space models and some of its typical methods to design feedback controllers include linear quadratic control, H-infinity control, and model predictive control~\cite{kalman1960contributions, zhou1998essentials, mayne2014model}. 
Since World War II, the advent of digital computers has led to significant advances in the application of modern control theory in areas including satellite navigation, rocket control, and autonomous driving.

A key feature of the aforementioned two control theories is that their stability analysis and controller synthesis are based on known system models. Take aircraft control as an example, one first uses the laws of rigid-body dynamics and aerodynamics to build an analytical model of the aircraft, which can then be used for computing characteristic roots and optimal feedback gain. Although this pattern enjoys great theoretical rigorousness from the enduring development of classical and modern control theories, its downside is noticeable in that it must rely on an explicit model which is generally required to be accurate. The need for explicit and accurate models highly restricts this control pattern on many complex systems whose dynamics are extremely difficult to model, such as soft robots, fluid systems, and stock markets.
Recent years have witnessed the rise of algorithm design and system development that are centralized in data. One hot topic in this field is deep learning, which uses neural networks to extract hierarchical features and learn complex patterns from a large quantity of labelled or unlabelled data. One may ask whether it is possible to combine data representation with control theories and construct a new control paradigm. The affirmative answer lies in the growing attention towards control methods based on data-type system description.
Some representative examples encompass iterative linear quadratic regulator~\cite{li2004iterative}, data-driven predictive control~\cite{berberich2020data}, and model-free reinforcement learning~\cite{guan2021direct}. One major difference between these new methods and traditional control methods is that their controller design does not use explicit models, and the system dynamics are described by state-action samples. Take model-free reinforcement learning as an example: it collects samples through repeating environment interaction, which serve as a data-type representation of environment dynamics. With sufficient data samples, a parameterized policy can be trained in an offline manner with mechanisms such as policy iteration or value iteration, and then be applied online for closed-loop control~\cite{li2023reinforcement}. 

\begin{figure*}
    \centering
    \includegraphics[width=0.7\linewidth]{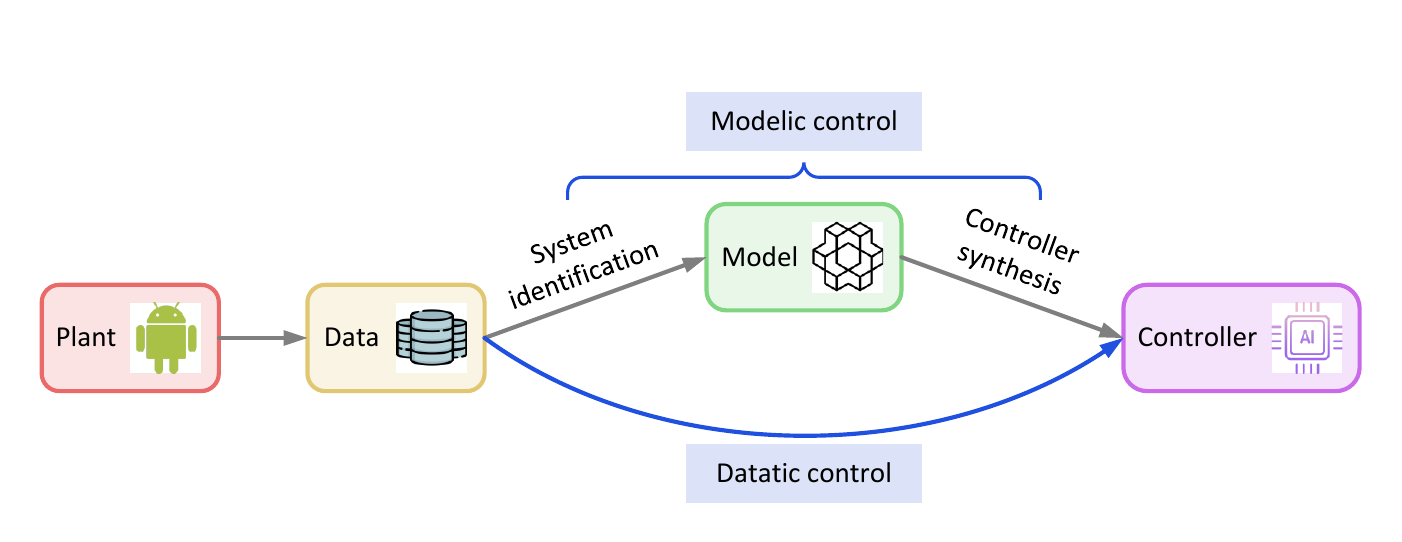}
    \caption{Two types of control paradigms. Modelic control (on the upper path) first performs system identification and then synthesizes controllers. Datatic control (on the lower path) directly solves controllers using data.}
    \label{fig: two control paradigms}
\end{figure*}

\begin{figure}
    \subfloat[Modelic description]{
        \includegraphics[width=0.5\linewidth]{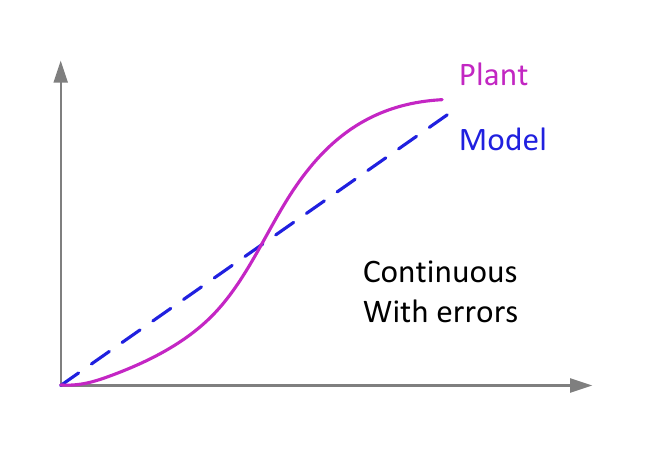}
    }
    \subfloat[Datatic description]{
        \includegraphics[width=0.5\linewidth]{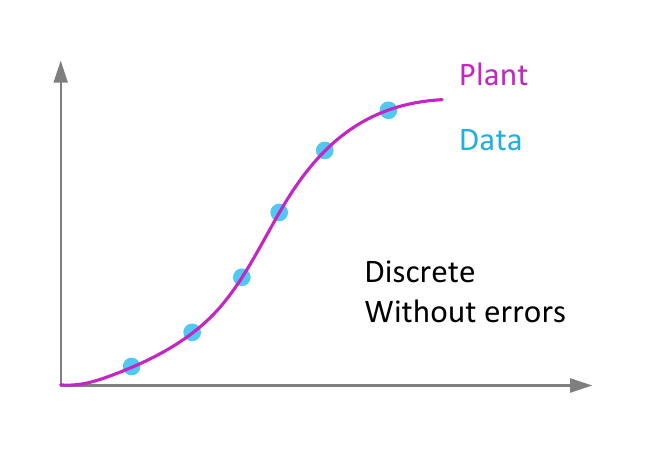}
    }
    \caption{System descriptions in two control paradigms. Modelic description is continuous but maybe erroneous. Datatic description is discrete but error-free.}
    \label{fig: system description}
\end{figure}

Through the analysis above, we can summarize two control paradigms according to whether the system dynamics is described by model or data: (1) \textit{modelic} control and (2) \textit{datatic} control, as shown in Figure \ref{fig: two control paradigms}. Here, ``modelic" and ``datatic" are two newly coined words, where ``modelic" means \textit{model-driven}, \textit{model-based} or \textit{model-related}, and ``datatic" has a corresponding meaning related to data. In modelic control paradigm, one first uses data to fit an analytical model through system identification and then uses this model to synthesize controllers. In contrast, datatic control paradigm directly solves controllers using data, eliminating the step of system identification. Obviously, both classical and modern control theories belong to the modelic control paradigm. As shown in Figure \ref{fig: system description}, whether an explicit model is used or not leads to different accuracy of system description. In a modelic control system, one needs to fit the system model with a function of a specific form. The model provides a continuous description of system dynamics, i.e., it can give an output at every point in the state-action space. However, modelic description is prone to errors because the true system may not exactly match the assumed function form. In a datatic control system, on the other hand, one does not build any explicit models but directly uses data samples to describe system dynamics. This kind of system behavior representation is called datatic description. Data samples, if sufficiently enough, can provide an accurate description of system dynamics (at least at their own locations) because they come from direct measurement of system states. One may question that if there are perception errors, datatic description will also be inaccurate. Actually, sensors can be viewed as part of a closed-loop system and thus their errors are also part of the system dynamics. Unfortunately, sensor measurement is not continuous in both temporal and spatial domains but only in the form of limited number of data points. There is no information in the interval of data points. Therefore, the datatic description of a dynamic system must be discrete, rather than continuous in the state-action space.

Stability has been a long-standing concern since early applications of feedback control such as James Watt's incorporation of centrifugal governor in the steam engine.
In the control community, there are various types of stability, including Lyapunov stability, orbital stability, and structural stability~\cite{pradeep1990stability}.
The most important type is Lyapunov stability, which states that for small values of initial disturbances, if the disturbed motion constrains itself to an arbitrarily prescribed small region of state space, then the system is said to be stable.
Early stability analysis methods trace back to the work of Routh~\cite{routh1877treatise} and Hurwitz~\cite{hurwitz1895ueber}, which verifies stability by examining the signs and magnitudes of coefficients of characteristic equations. Later, Nyquist criterion~\cite{nyquist1932regeneration} lays the foundation for stability analysis in the frequency domain, which uses the frequency response of open-loop function to predict the closed-loop stability. Another frequency response tool is the root locus method introduced by Evans~\cite{evans1948graphical}, which plots the closed-loop poles of the system as a function of loop gain. The most popular stability theory in control might be Lyapunov stability criterion~\cite{lyapunov1992general}, which infers the stability of equilibrium points by constructing a Lyapunov function that satisfies certain properties. Other representative works on stability theory include Bogoliubov's integral manifolds method~\cite{bogoliubov1961asymptotic}, Popov's conditions for absolute stability~\cite{popov1961absolute}, and functional analytic approach by Sandberg and Zames~\cite{sandberg1965some, zames1966input}.
It is easy to find that the abovementioned stability criteria all rely on system models of some kinds. We know that the difference between modelic description and datatic description is that the former provides system information in the whole space while the latter only provides information on data points but not the region between data points (see Figure \ref{fig: system description}(b)). This makes stability unable to be rigorously tested beyond data points in those systems with datatic description. Therefore, we need to develop a new stability criterion for datatic control systems that can break through the barricade of system description being intermittently absent over the state-action space. Such a criterion will suit the verification of \textit{datatic stability} because it solemnly depends on data samples rather than a system model.

Over the past several years, there have been some exploratory works about the datatic stability of linear systems, also known as data-driven stability verification.
Most of these works are based on Willems et al.'s fundamental lemma~\cite{willems2005note}, which shows that persistently exciting data can be used to represent the input-output behavior of a linear system.
Persis et al. (2019) parameterize linear systems with state-action pairs and solve linear controllers using data-dependent linear matrix inequalities~\cite{de2019formulas}. The control problems they solve include output feedback stabilization and linear quadratic regulation.
Waarde et al. (2020) extend Willems' lemma to the situation where multiple system trajectories are given instead of a single one~\cite{van2020willems}. They introduce a notion of collective persistency of excitation and show that a linear system can be described by a finite number of state-action trajectories.
Waarde and his colleagues further propose the concept of data informativity to deal with trajectories that are not persistently exciting~\cite{van2020data}. They investigate necessary and sufficient conditions on the informativity of data for several datatic control problems, such as stability, controllability, and stabilization.
The reason why the datatic stability of linear control systems can be tested is that there is a strong assumption, i.e., data points and their intervals are linearly continuous. This allows model information to be easily extended from data points to data-absent regions. In nonlinear control systems, however, such a linear continuity no longer holds and all the above methods become invalid.

Recently, some researchers begin to study the datatic stability of nonlinear systems with special characteristics. Although the system dynamics are no longer required to be linear, they are still assumed to be in some fixed functional forms, which enable continuous extension from known data points to unknown regions. 
For example, Lavaei et al. (2022) explore the datatic stability of homogeneous control systems of degree one~\cite{lavaei2022data}. The homogeneity of system dynamics ensures that there exists a homogeneous Lyapunov function and when constructing such a function, the Lyapunov conditions only need to be checked on a unit ball instead of the whole state space. Using these properties, stability verification is cast as robust optimization problem, whose approximate solution is given with a certain confidence level.
Guo et al. (2021) deal with polynomial control systems with known maximum degree~\cite{guo2021data}. The dynamics of such a system is written as a linear function of monomial vectors and its Lyapunov function is constructed as a quadratic form of a monomial vector. Therefore, its stability verification is equivalent to finding a positive definite matrix in the Lyapunov function, which is a sum-of-squares program.
Choi et al. (2021) discuss a control-affine nonlinear system with two polynomial approximations~\cite{choi2021convex}. They first approximate the stability certificate with polynomials, which allows controller design to be a sum-of-squares program. Then they approximate two operators related to system dynamics in the program with polynomial basis functions, which enables datatic stability to be verified by a data-driven optimization algorithm.
The aforementioned three works focus on nonlinear systems with specific forms, e.g., homogeneous systems, polynomial systems, or systems with polynomial approximation. In essence, these forms make certain assumptions about the inherent connection among data points. For example, polynomial systems assume that information between data points can be accurately described by polynomials, which all follow the same specific structure, significantly narrowing down the occurrence possibilities of abnormal system behaviors in unmeasured regions. Although these assumptions are weaker than linearity, they still cannot be satisfied in general nonlinear systems where regions between data points may be connected by various functions. As far as we know, a universal stability criterion for nonlinear datatic control systems is still missing.

\begin{figure}
    \centering
    \subfloat[Linear system dynamics]{
        \includegraphics[width=0.47\linewidth, trim={0 2450 0 0}, clip]{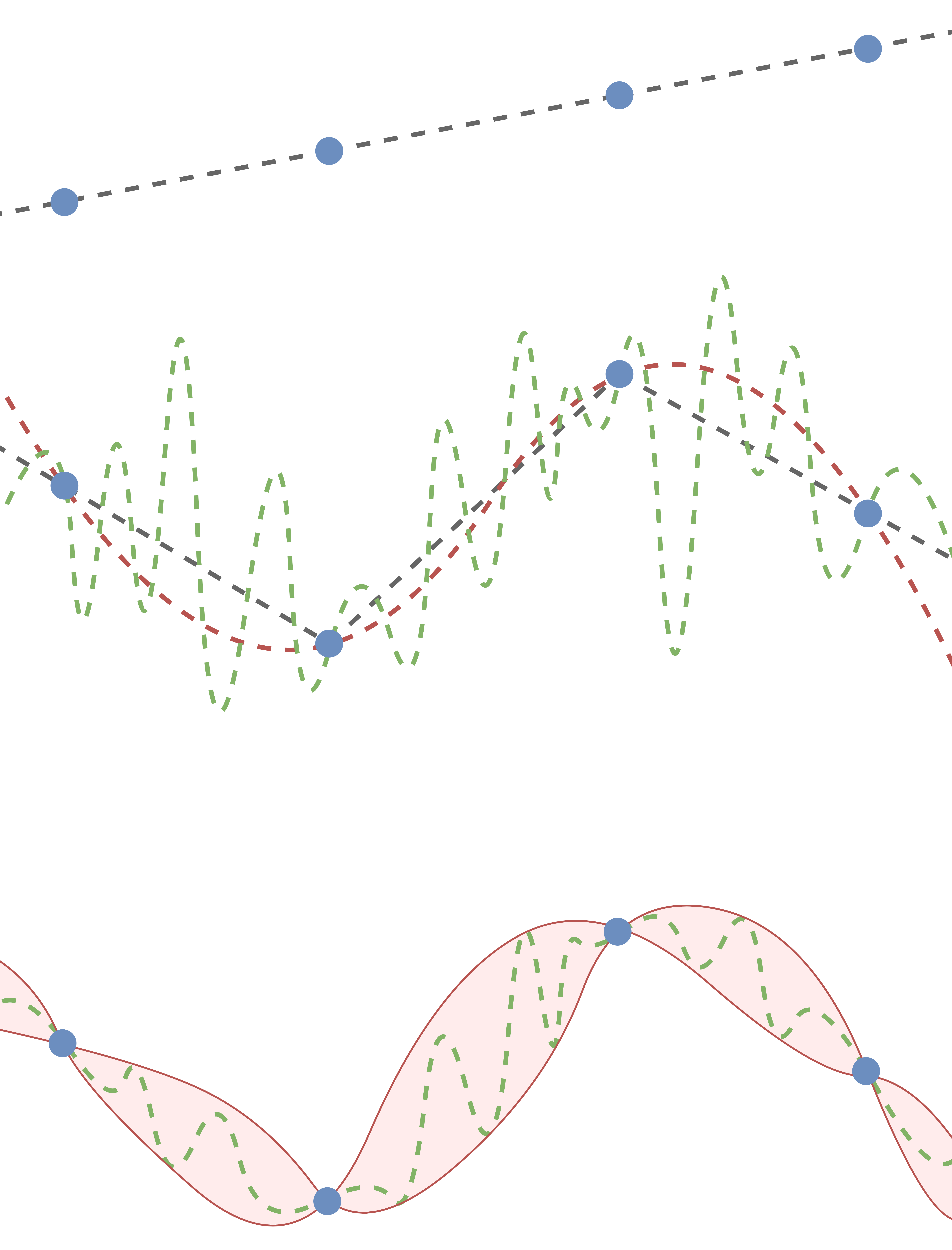}
    }
    \subfloat[Nonlinear system dynamics]{
        \includegraphics[width=0.47\linewidth, trim={0 1250 0 650}, clip]{figure/f_estimation.png}
    }
    \\
    \subfloat[Continuous extension of nonlinear system dynamics]{
        \label{fig: system dynamics schematic-c}\includegraphics[width=0.6\linewidth, trim={0 0 0 1900}, clip]{figure/f_estimation.png}
    }
    \caption{Schematics of different system dynamics. The dashed lines stand for possible system dynamics. The blue circles represent data points where the true values of dynamics are measured. For a linear system, the value of system dynamics between data points can be easily determined given sufficient linearly independent points, while this does not hold for a nonlinear system. Nevertheless, with the assumption of continuity, we are able to extend system information from data points to unknown regions, leading to an estimation of the range of system dynamics (see the red shaded region in Fig.\ref{fig: system dynamics schematic-c}).}
    \label{fig: system dynamics schematic}
\end{figure}

% Objective
This paper for the first time introduces a systematic stability verification method for general datatic control systems. More specifically, given a control policy, we can check its stability in a system described by a set of input and output data, without needing to explicitly build the dynamic model.
% Method
The stability verification is based on how to estimate the time derivative of Lyapunov function. As we have discussed before, a datatic control system suffers from the information absence in unmeasured regions. To conquer this issue, we assume that the system dynamics is Lipschitz continuous and formulate the stability verification problem as a quadratically constrained linear program (QCLP). Our algorithm is tested on both linear and nonlinear datatic control systems. Results show that our algorithm successfully verifies the cases of stability, instability, and critical stability. The main contributions of this paper are summarized as follows.
\begin{enumerate}
    \item We leverage the Lipschitz continuity assumption of system dynamics to perform a continuous extension from data points to state-action pairs outside the dataset. We find that the time derivative of a system state corresponding to each state-action pair lies in the intersection of a set of closed balls given by data points. The radius of balls depends on local Lipschitz constants of system dynamics, which are estimated by solving a quadratic program.
    \item We test the stability of a nonlinear datatic control system by solving a set of QCLPs. The objective of QCLP is to maximize the time derivative of Lyapunov function, and its constraints are the intersection of closed balls governed by the Lipschitz continuity of system dynamics. According to Lyapunov's second method, stability relies on whether the solutions of QCLPs are less than zero. Therefore, QCLP is solved at each state point to check whether all maximum values are less than zero.
\end{enumerate}

\section{Preliminaries}
\subsection{Datatic control systems}
A standard datatic control system includes a set of input and output data collected by interacting with a plant, which is denoted as
\begin{equation}
\label{eq: data}
    \mathcal{D}=\{(x_i,u_i,\dot{x}_i)|1\le i\le N\},
\end{equation}
where $x\in\calX\subseteq\mathbb{R}^n$ is the state, $u\in\calU\subseteq\mathbb{R}^m$ is the action, $\dot{x}$ is the time derivative of state, and $N$ is the number of data items.
The dataset $\mathcal{D}$ is a datatic description of a continuous-time plant:
\begin{equation}
\label{eq: dynamics}
    \dot{x}=f(x,u),
\end{equation}
where $f:\calX\times\calU\to\calX$ is an unknown continuous function. That is to say,
\begin{equation}
    \dot{x}_i=f(x_i,u_i), \forall i=1,2,\dots,N.
\end{equation}
Equation \eqref{eq: dynamics} is actually a modelic description of plant dynamics. If it is known and accurate, we can use it to verify system stability. Unfortunately, models are usually inaccurate or even unknown in many real-world tasks. How to directly use data in \eqref{eq: data} for stability verification is the focus of this paper.

\subsection{Reshape Lyapunov stability criterion}
Lyapunov stability describes the ability of an autonomous system to stay in the vicinity of an equilibrium point. Given a policy $\pi:\calX\to\calU$, the original control system \eqref{eq: dynamics} turns into an autonomous system:
\begin{equation}
\label{eq: autonomous system}
    \dot{x}=f(x,\pi(x)).
\end{equation}
Its equilibrium $x_e$ is said to be Lyapunov stable if the system state can stay within an arbitrarily small neighborhood of $x_e$ forever as long as the initial state is close enough to $x_e$. 
Without loss of generality, we assume \eqref{eq: autonomous system} has only one equilibrium point at $x_e=0$. To check its stability, we can use Lyapunov's second method described as follows.

\begin{theorem}
\label{thm: Lyapunov stability criterion}
For an autonomous system \eqref{eq: autonomous system}, if there exists a continuously differentiable function $V:\calX\to\mathbb{R}$, such that
\begin{enumerate}
    \item $V(x)\ge0,\forall x\in\calX$ and $V(x)=0$ iff $x=0$.
    \item $\dot{V}(x)\le0,\forall x\in\calX$ and $\dot{V}(x)=0$ iff $x=0$.
\end{enumerate}
Then $V$ is called a Lyapunov function and the system is stable in the sense of Lyapunov.
\end{theorem}

Given a datatic description $\mathcal{D}$, a policy $\pi$, and a function $V$, our goal is to verify whether the autonomous system yielded by $\pi$ is Lyapunov stable. More specifically, we seek to check whether the two conditions in Theorem \ref{thm: Lyapunov stability criterion} are fulfilled on all state points in $\mathcal{D}$ under the policy $\pi$. Since the Lyapunov function $V$ is known, it is easy to check the first condition. Our main concern is to check the second condition, i.e., whether
\begin{equation}
    \label{equ:Lyapunov second condition}
    \dot{V}(x_i)=\dv{V(x)}{x^\top}\bigg|_{x=x_i}f(x_i,\pi(x_i))\le0
\end{equation}
holds for all $i=1,2,\dots,N$. Since data collection and stability verification are two different tasks, the policy to be verified is often different from the one used to collect data, i.e., $u_i\neq\pi(x_i)$, where $(x_i,u_i)\in\mathcal{D}$. Therefore, the point $(x_i,\pi(x_i))$ is not in the dataset in most cases, and the value of $f$ on this point is unknown, which makes it difficult to examine condition \eqref{equ:Lyapunov second condition} in a datatic control system.

\section{Stability Criterion of Datatic Control Systems}
The core idea of our stability criterion is to extend the information of plant dynamics on known data points to unknown regions. The extension result is a range estimation of plant dynamics on unknown state-action pairs. The extension is based on the Lipschitz continuity assumption of plant dynamics, which restricts the value of $f$ on unknown points to a certain range. In this restricted range, we can compute the worst-case value of the left-hand side of \eqref{equ:Lyapunov second condition} and check whether it is below zero.

\subsection{Continuity assumption of plant dynamics}
The crux of examining condition \eqref{equ:Lyapunov second condition} is that all information we know about $f$ is restricted to a limited number of points $\sbr{x_i,u_i}$, yet we attempt to evaluate it on other different points $\sbr{x_i,u^\pi_i}$, where $u^\pi_i=\pi(x_i)$. This makes a continuity assumption on the plant dynamics inevitable.

The condition that $f$ is continuous alone is not enough to extend the information of $u_i$ to $u^\pi_i$ because the derivatives of each order are not specified. This may cause the value of $f$ at $\sbr{x_i,u^\pi_i}$ to be significantly different from that at $\sbr{x_i,u_i}$ even if it is very close to the latter point. In this regard, a feasible method is to force the derivative of $f$ to be also continuous, i.e., $f$ is continuously differentiable. Accordingly, its derivative is bounded on a bounded closed set, and the rate of change becomes limited. In many cases, the assumption of continuous differentiability is too strong since it requires the function to be differentiable. Here, we use the Lipschitz continuity instead, which is a necessary condition for continuous differentiability. The good side of this replacement is that Lipschitz continuity does not require the differentiability of $f$.

\begin{assumption}[Lipschitz continuity]
    \label{asmp:Lipschitz continuity}
    $f(x,u)$ is Lipschitz continuous with respect to $x$ and $u$, i.e., 
    \begin{equation}
        \begin{aligned}
            \forall (x_1,u&), \sbr{x_2,u}\in\calX\times\calU, \\
            &d_{\calX}\sbr{f\sbr{x_1,u}, f\sbr{x_2,u}} \le L_x d_{\calX}\sbr{x_1, x_2}, \\
            \forall (x,u_1&), \sbr{x,u_2}\in\calX\times\calU, \\
            &d_{\calX}\sbr{f\sbr{x,u_1}, f\sbr{x,u_2}} \le L_u d_{\calU}\sbr{u_1,u_2},
        \end{aligned}
    \end{equation}
    where $d_\calX(\cdot,\cdot)$ and $d_{\calU}(\cdot,\cdot)$ are some metrics on $\calX$ and $\calU$, and $L_x$ and $L_u$ are Lipschitz constants with respect to $x$ and $u$.
\end{assumption}

Generally, $d_\calX(\cdot,\cdot)$ and $d_{\calU}(\cdot,\cdot)$ are chosen to be the Euclidean distance, and they are abbreviated as $d(\cdot,\cdot)$ in the rest of the paper.

\subsection{Stability verification}
\begin{figure}
    \centering
    \includegraphics[width=0.8\linewidth]{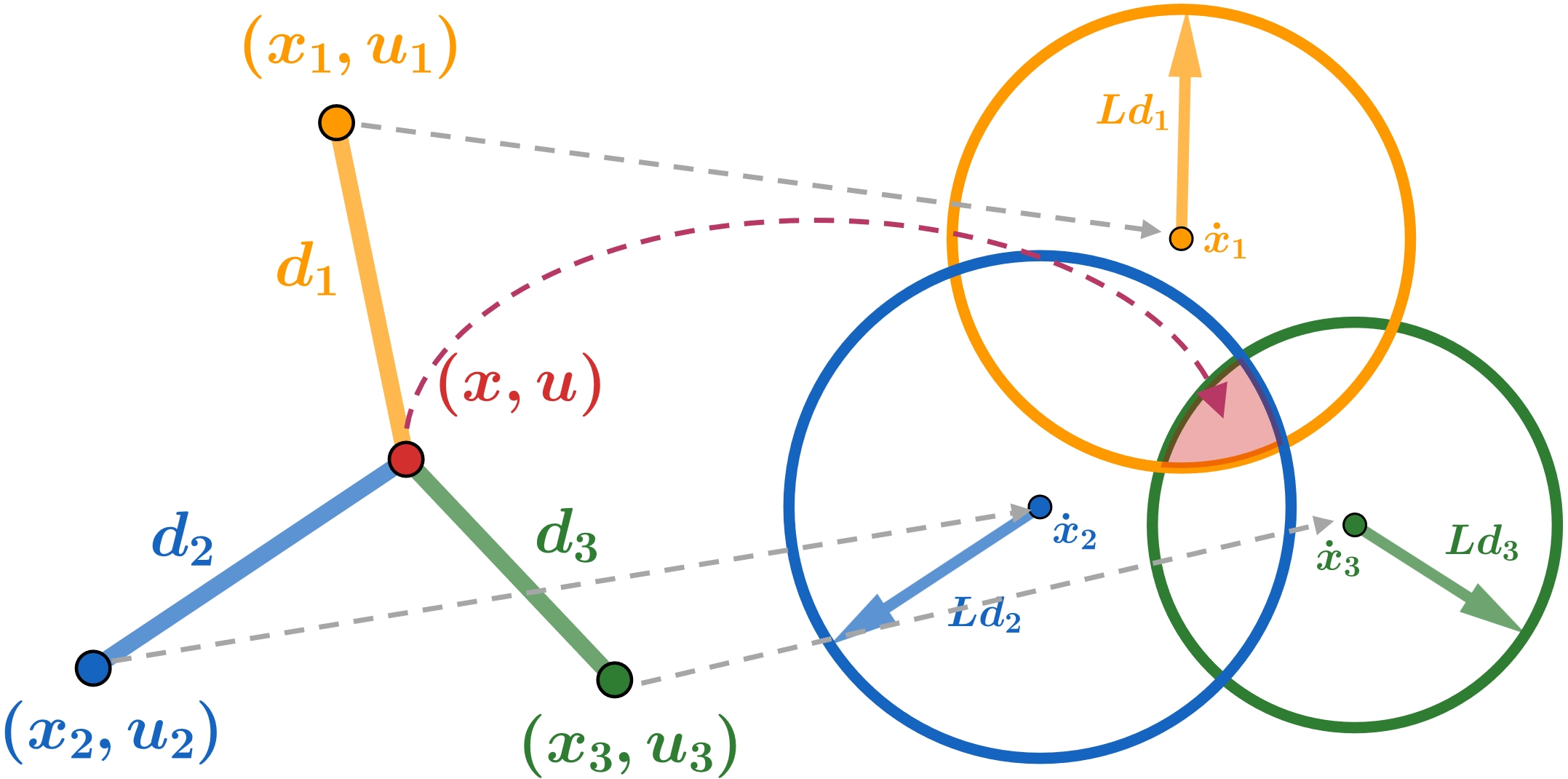}
    \caption{Intersection of closed balls given by Lipschitz continuity.}
    \label{fig: Lipschitz continuity}
\end{figure}

The inequality to be inspected in \eqref{equ:Lyapunov second condition} involves two parts, $\mathrm{d}V/\mathrm{d}x$ and $f(x,\pi(x))$, the former of which is at hand while the latter is not. Based on Assumption \ref{asmp:Lipschitz continuity}, we can make a continuous extension on $f$, from the known data points to unknown regions, so that we can restrict the value of $f$ on an arbitrary point to a certain range.
For simplicity of notation, we denote
$$\dot{x}^\pi_i\triangleq f(x_i,\pi(x_i)), i=1,2,\dots,N.$$
The key is to find a range estimate of $\dot{x}^\pi_i$. The more accurate the estimate is, the better it is for stability verification. Consider its relationship with the $j$-th data point $(x_j,u_j,\dot{x}_j)$. Using triangular inequality and Lipschitz continuity, we have
\begin{equation}
    \begin{aligned}
         d\sbr{\dot{x}^\pi_i, \dot{x}_j}&\le d\sbr{\dot{x}^\pi_i,f(x_j,u^\pi_i)}+d\sbr{f(x_j,u^\pi_i),\dot{x}_j} \\
         &\le L_x d\sbr{x_i,x_j} + L_u d\sbr{u^\pi_i, u_j}.
    \end{aligned}
\end{equation}
This is equivalent to
$$
\dot{x}^\pi_i \in \mathcal{B}\sbr{\dot{x}_j,r_{ij}},
$$
where
$$
\begin{aligned}
    \mathcal{B}\sbr{\dot{x}_j,r_{ij}}=\bbr{\dot{x}|d(\dot{x}_j,\dot{x})\le r_{ij}}, \\
    r_{ij}=L_x d\sbr{x_i,x_j} + L_u d\sbr{u^\pi_i, u_j}.
\end{aligned}
$$
That is to say, $\dot{x}^\pi_i$ lies within a closed ball centered at $\dot{x}_j$ with a radius of $r_{ij}$. For each data point $\sbr{x_j,u_j,\dot{x}_j}$, we can construct such a closed ball where $\dot{x}^\pi_i$ must lie within. The intersection of all these $N$ balls is the best estimation we can have on $\dot{x}^\pi_i$ in the sense that it includes the least candidates of $\dot{x}^\pi_i$ by considering all data points:
\begin{equation}
\label{eq: intersection of balls}
    \dot{x}^\pi_i \in \bigcap_{j=1}^N\mathcal{B}\sbr{\dot{x}_j, r_{ij}}.
\end{equation}
Figure \ref{fig: Lipschitz continuity} gives an illustration of \eqref{eq: intersection of balls}. The value of $f$ at point $(x,u)$ is to be estimated. Three data points $(x_1,u_1)$, $(x_2,u_2)$, and $(x_3,u_3)$ can provide information for this estimation. Using Lipschitz continuity, each data point gives a closed ball that $f(x,u)$ must lie within. The intersection of closed balls (red region) is the best estimation we can obtain given these three data points.

\begin{figure}
    \centering
    \includegraphics[width=0.8\linewidth]{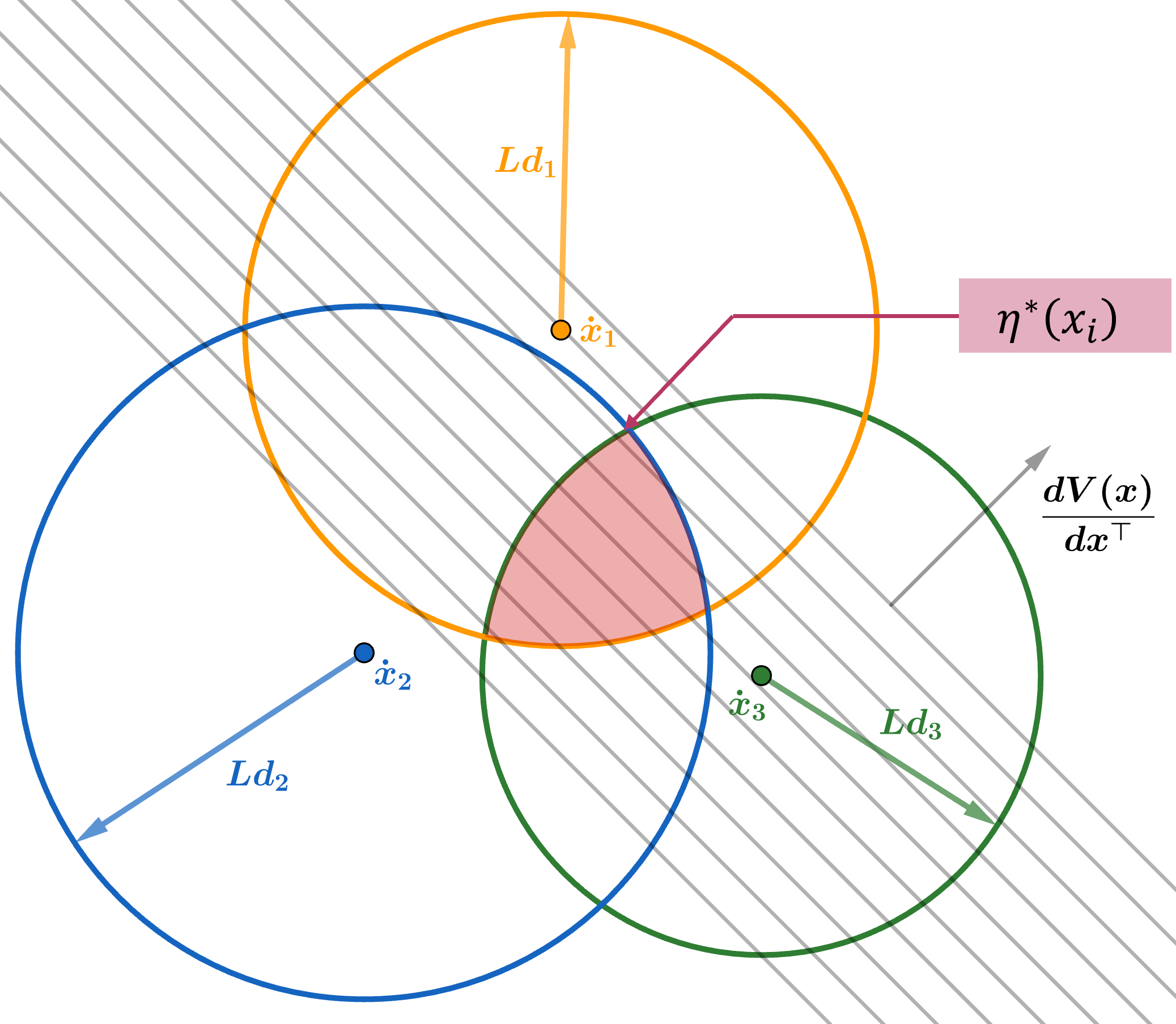}
    \caption{QCLP for finding the maximum value of time derivative of Lyapunov function.}
    \label{fig: QCLP}
\end{figure}

Since every value in the intersection is a possible candidate of the true $\dot{x}^\pi_i$, the problem of checking stability then becomes finding the maximum time derivative of $V(x)$ over the intersection. Since $\dot{V}(x)$ is a linear function of $\dot{x}^\pi_i$ (see \eqref{equ:Lyapunov second condition}) and the constraint of each closed ball can be written in a quadratic form, the datatic stability verification problem can be formulated as a quadratically constrained linear program (QCLP):
\begin{equation}
\label{equ:QCLP}
\begin{aligned}
    \max_{\dot{x}} \quad & \eta=\dv{V(x)}{x^\top}\bigg|_{x=x_i}\dot{x} \\
    \st \quad & (\dot{x}-\dot{x}_j)^\top(\dot{x}-\dot{x}_j)\le r_{ij}^2, \forall j=1,2,\dots,N.
\end{aligned}
\end{equation}
We denote the objective function of \eqref{equ:QCLP} as $\eta$ to distinguish it from the true time derivative of Lyapunov function $\dot{V}(x_i)$. The symbol $\eta$ is called \textit{stability index}. The optimal value of QCLP is denoted as $\eta^*$. Note that $\eta^*$ is related to the state point $x_i$, i.e., each state point has its own optimal value.
Figure \ref{fig: QCLP} gives an illustration of QCLP. Given a state point $x_i$, $\mathrm{d}V(x)/\mathrm{d}x^\top\big|_{x=x_i}$ is a known, deterministic vector, which corresponds to the normal vector of parallel lines in Figure \ref{fig: QCLP}. The red region corresponds to all possible values of $\dot{x}$. The further $\dot{x}$ goes to the upper right part of this region (in the direction of normal vector), the larger its inner product with the normal vector becomes. Therefore, the optimal value of QCLP is obtained when $\dot{x}$ is at the upper right corner. 
Intuitively, such a QCLP considers all plant dynamics compatible with both Lipschitz continuity and collected data. If the optimal value $\eta^*$ is less than zero, then the true time derivative $\dot{V}(x_i)$ must also be less than zero. By solving such a QCLP on each state point and comparing the optimal value with zero, we know for sure whether the second Lyapunov condition is fulfilled on all state points in $\mathcal{D}$. This method for stability verification of datatic control system is called \textit{$\eta$-testing}. The outcomes of $\eta$-testing is summarized as follows:
\begin{equation}
\label{eq: eta-testing}
    \begin{aligned}
        \forall x_i\neq0\in\mathcal{D},\eta^*(x_i)<0 \implies &\text{\eqref{eq: autonomous system} is Lyapunov stable.} \\
        \exists x_i\neq0\in\mathcal{D},\eta^*(x_i)\ge0 \implies &\text{Lyapunov stability of \eqref{eq: autonomous system}} \\
        &\text{is undeterminable.} 
    \end{aligned}
\end{equation}

% √TODO: 加一节，说明如果在所有点上都检验了不等式，那么就是系统稳定的充分条件。但是实际不能检验所有点，只能在数据集上检验。
% 画包含关系图：系统稳定、Lyapunov条件、\eta-testing

\begin{figure}
    \centering
    \includegraphics[width=0.6\linewidth]{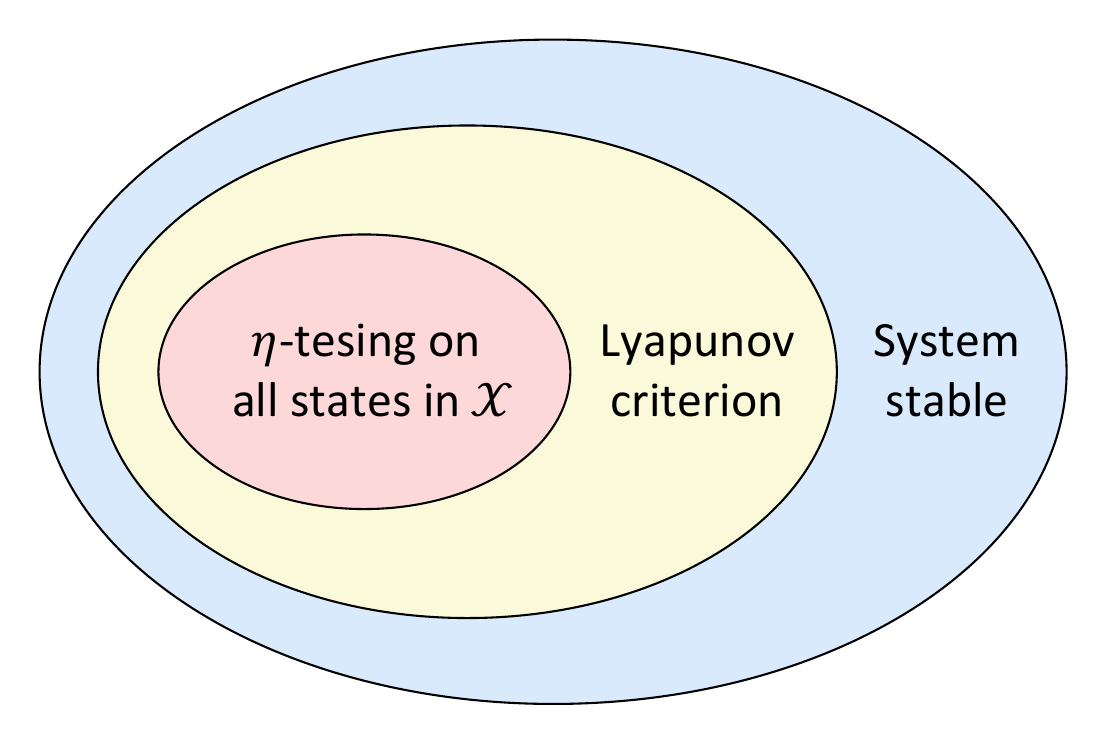}
    \caption{Containment relationship of $\eta$-testing on all states in state space $\mathcal{X}$, Lyapunov criterion, and system stability.}
    \label{fig: containment}
\end{figure}

% √TODO: 明确写出公式10的不准确，体现D和X的对比
In fact, \eqref{eq: eta-testing} is not a rigorous stability criterion because it only verifies Lyapunov condition on states in the dataset $\mathcal{D}$ instead of the whole state space $\mathcal{X}$. The state space $\mathcal{X}$ is a continuous set in $\mathbb{R}^n$ with infinite states while the dataset $\mathcal{D}$ is only a discrete set with a finite number of states. Even if the Lyapunov condition is verified on all states in $\mathcal{D}$, the system still may not be stable because the condition may be violated on some states outside $\mathcal{D}$ but in $\mathcal{X}$. To rigorously verify system stability, one must perform $\eta$-testing on all states in $\mathcal{X}$. If this test is possible, it provides a sufficient condition for Lyapunov criterion, which in turn is a sufficient condition for system stability, as shown in Figure \ref{fig: containment}. However, performing $\eta$-testing on all states in $\mathcal{X}$ is impractical because it requires solving an infinite number of QCLPs. In practice, we only perform it on the dataset $\mathcal{D}$, resulting in an approximate criterion for system stability. So long as one can collect enough data points, the stability verification will become sufficiently accurate.
% In theory, if QCLP \eqref{equ:QCLP} is solved on all states in the state space, and all optimal values are less than zero, then a sufficient condition for stability of system \eqref{eq: autonomous system} is achieved. To see this, we note that for any state $x$, the optimal value of its corresponding QCLP is the worst-case estimate of $\dot{V}(x)$ considering all possible values of $\dot{x}$. Therefore, if the optimal value of QCLP is less than zero, the true value of $\dot{V}(x)$ must also be less than zero. This explains why performing $\eta$-testing on all states yields a sufficient condition for $\dot{V}(x)<0$ in Lyapunov criterion. Since Lyapunov criterion is itself a sufficient condition for system stability, we conclude that $\eta$-testing also gives a sufficient condition. The containment relationship of $\eta$-testing, Lyapunov criterion, and system stability is shown in Figure \ref{fig: containment}. 

Next, we discuss some variants of $\eta$-testing, including how to verify unstable systems and how to handle discrete-time systems.

\subsection{Instability verification}
As shown above, if the optimal solution of \eqref{equ:QCLP} is greater than or equal to zero on any data point, system stability is undeterminable. This is because Lyapunov's second method itself is only a sufficient condition for stability. Moreover, we can only make a worst-case estimate of time derivative of Lyapunov function, which further makes our stability criterion a sufficient condition.
To check instability, we need to use a symmetrical variant of Lyapunov’s second method expressed as follows.

\begin{theorem}
\label{thm: Lyapunov instability criterion}
For an autonomous system \eqref{eq: autonomous system}, if there exists a Lyapunov function $V:\calX\to\mathbb{R}$, such that
$$\dot{V}(x)\ge0,\forall x\in\calX \text{ and } \dot{V}(x)=0 \text{ iff } x=0.$$
Then the system is unstable in the sense of Lyapunov.
\end{theorem}

The difference between Theorem \ref{thm: Lyapunov instability criterion} and Theorem \ref{thm: Lyapunov stability criterion} is that the sign of $\dot{V}$ is reversed. As a result, we need to compute the minimum value of $\dot{V}$ instead of its maximum and compare it with zero to verify instability. Since the Lipchitz continuity assumption still holds, the optimization variable $\dot{x}$ lies in the intersection of a set of closed balls. Therefore, the instability verification problem can also be formulated as a QCLP, with the maximization in \eqref{equ:QCLP} changed to minimization:
\begin{equation}
\label{eq: QCLP for instability}
    \min_{\dot{x}} \quad \eta
\end{equation}
Here, its constraints are the same as those in \eqref{equ:QCLP}. We solve problem \eqref{eq: QCLP for instability} on each state point in the dataset and compare the minimum values $\eta^*$ with zero. If the minimum values on all state points are above zero, the system is unstable.

\subsection{Stability of discrete-time systems}
The aforementioned stability criterion is for continuous-time systems and it can be extended to discrete-time systems as well. In a discrete-time system, the plant dynamics takes the form of
\begin{equation}
    \label{equ:discrete-time system}
    x^\prime=f(x,u), f:\calX\times\calU\to\calX,
\end{equation}
where $x^\prime$ is the next state. The datatic description of this plant is thereby
\begin{equation}
    \mathcal{D}=\{(x_i,u_i,x^\prime_i)|1\le i\le N\}.
\end{equation}
Lyapunov's second method for the stability of discrete-time systems is slightly different from that of continuous-time systems.
% √TODO: differs a little 换一个词
\begin{theorem}
\label{Lyapunov's second method for stability discrete}
For a discrete-time autonomous system, if there exists a Lyapunov function $V:\calX\to\mathbb{R}$ such that
$$V(x^\prime)-V(x)\le0,\forall x\in\calX \text{ and } V(x^\prime)-V(x)=0 \text{ iff } x=0,$$
the system is stable in the sense of Lyapunov.
\end{theorem}

Similarly, for each state $x_i$, we need to check whether the maximum possible value of $V(x_i^{\pi\prime})-V(x_i)$ is less than zero. This turns out to be such an optimization problem:

\begin{equation}
\label{equ:QCLP discrete}
\begin{aligned}
    \max_{x^\prime} \quad & V(x^\prime) \\
    \st \quad & (x^\prime-x^\prime_j)^\top(x^\prime-x^\prime_j)\le r_{ij}^2, \forall j=1,2,\dots,N.
\end{aligned}
\end{equation}

\section{Stability Criterion of Linear Datatic Control Systems}
In this section, let us turn to the stability verification of linear systems. As introduced above, there have been some existing works in this topic~\cite{de2019formulas, van2020data}. Here, we will talk about how our method is simplified for linear systems, as well as its connection with existing studies.

\subsection{Criterion for continuous-time linear systems}
For a datatic control system, linearity is more informative than nonlinearity. Linearity means that the plant dynamics between data points can be described by a linear function, while nonlinearity cannot give accurate information about the dynamics in data-absent regions, as shown in Figure \ref{fig: system dynamics schematic}. Therefore, the continuous extension is indispensable in nonlinear systems but can be largely simplified in linear systems. With the knowledge of linear connection, our stability criterion changes from solving optimization problems to checking the negative definiteness of a matrix. The dynamics of a linear system is
\begin{equation}
\label{eq: linear dynamics}
    \dot{x}=Ax+Bu,
\end{equation}
where $A\in\mathbb{R}^{n\times n}$ is the system matrix and $B\in\mathbb{R}^{n\times m}$ is the control matrix. Both $A$ and $B$ are unknown. The form of input and output data is the same as \eqref{eq: data}. Since data obeys the plant dynamics, it can be written in a compact form:
\begin{equation}
\label{eq: linear data}
    \dot{X}=AX+BU,
\end{equation}
where
$$
\begin{aligned}
    \dot{X}&=
    \begin{bmatrix}
        \dot{x}_1 & \dot{x}_2 & \dots & \dot{x}_N
    \end{bmatrix}, \\
    X&=
    \begin{bmatrix}
        x_1 & x_2 & \dots & x_N
    \end{bmatrix}, \\
    U&=
    \begin{bmatrix}
        u_1 & u_2 & \dots & u_N
    \end{bmatrix}.
\end{aligned}
$$
The control policy to be checked is a state feedback law:
\begin{equation}
\label{eq: linear policy}
    u=Kx,
\end{equation}
where $K\in\mathbb{R}^{m\times n}$ is the feedback gain. The closed-loop dynamics under this linear policy is
\begin{equation}
    \dot{x}=(A+BK)x.
\end{equation}

In linear systems, Lyapunov functions are usually selected to be in a quadratic form:
\begin{equation}
\label{eq: quadratic Lyapunov}
    V(x)=x^\top Px,
\end{equation}
where $P\in\mathbb{R}^{n\times n}$ is a given positive definite matrix. A Lyapunov function of this form naturally satisfies the first condition in Theorem \ref{thm: Lyapunov stability criterion}. To verify the second condition, we first compute the time derivative of Lyapunov function:
\begin{equation}
\label{eq: Lyapunov derivative}
    \dot{V}(x)=x^\top(P(A+BK)+(A+BK)^\top P)x.
\end{equation}

To verify whether $\dot{V}(x)<0$ holds is equivalent to verify
\begin{equation}
\label{eq: Lyapunov derivative matrix}
    Q\triangleq P(A+BK)+(A+BK)^\top P\prec0,
\end{equation}
where $Q$ is called \textit{dissipation matrix}. The difficulty of this verification is that both $A$ and $B$ are unknown. To solve this problem, we need to replace these two matrices with data. Equation \eqref{eq: linear data} is equivalent to
$$
\dot{X}=
\begin{bmatrix}
    A & B
\end{bmatrix}
Z,
% \begin{bmatrix}
%     X \\ U
% \end{bmatrix}.
$$
where $Z$ is state-action matrix:
\begin{equation}
\label{eq: Z matrix}
    Z\triangleq
    \begin{bmatrix}
        X \\ U
    \end{bmatrix}.
\end{equation}
In order to guarantee the existence of the right inverse of $Z$, we require that it
has full row rank, i.e.,
\begin{equation}
\label{eq: full row rank}
    \mathrm{rank}(Z)=n+m.
\end{equation}

Existing works have some discussion of \eqref{eq: full row rank} from the perspective of persistency of excitation. Willems et al.~\cite{willems2005note} prove that a sufficient condition for \eqref{eq: full row rank} is that the action data is a persistently exciting sequence of order $n+1$. Waarde et al.~\cite{van2020willems} propose the concept of collective persistency of excitation, which extends Willems' result to the case of multiple sequences with lower excitation orders. Note that these results are only sufficient conditions for \eqref{eq: full row rank}. In fact, the action data is not necessarily in the form of sequences and can be just composed of independent data points. In this case, the action is not persistently exciting but we can still choose proper data points so that \eqref{eq: full row rank} is satisfied.
If \eqref{eq: full row rank} is satisfied, $Z$ has a right inverse, i.e., there exists $Z^+\in\mathbb{R}^{N\times(n+m)}$, such that
$$ZZ^+=I.$$
Thus, we can replace matrices $A$ and $B$ with data:
\begin{equation}
\label{eq: linear system representation}
\begin{bmatrix}
    A & B
\end{bmatrix}
=\dot{X}Z^+,
\end{equation}
and we have
$$
A+BK=
\begin{bmatrix}
    A & B
\end{bmatrix}
\begin{bmatrix}
    I \\ K
\end{bmatrix}
=\dot{X}Z^+
\begin{bmatrix}
    I \\ K
\end{bmatrix}.
$$
Therefore, the dissipation matrix is computed as
\begin{equation}
    Q=P\dot{X}Z^+
    \begin{bmatrix}
        I \\ K
    \end{bmatrix}+
    \left(
    \dot{X}Z^+
    \begin{bmatrix}
        I \\ K
    \end{bmatrix}
    \right)^\top P.
\end{equation}
If $Q$ is negative definite, the linear system is stable.

Through the above analysis, we can see that stability verification of linear systems does not require solving QCLPs as in nonlinear systems. Instead, stability can be verified by checking the negative definiteness of dissipation matrix. This is because in a linear system we can obtain the analytical expression of time derivative of Lyapunov function \eqref{eq: Lyapunov derivative}. Moreover, $A$ and $B$ can be replaced with data as long as the full row rank requirement \eqref{eq: full row rank} is satisfied. This requirement is easy to satisfy in practice because the number of data points is usually much larger than state and action dimensions, i.e., $N\gg n+m$. As long as there are $n+m$ linearly independent data points in the dataset, the full row rank requirement is satisfied.

\subsection{Criterion for discrete-time linear systems}
Similar to the case of nonlinear systems, our stability verification method suits for both continuous-time and discrete-time linear systems. The dynamics of a discrete-time linear control system is
\begin{equation}
\label{eq: discrete linear dynamics}
    x'=Ax+Bu,
\end{equation}
Its compact form with collected data is
\begin{equation}
\label{eq: discrete linear data}
    X'=AX+BU,
\end{equation}
where
$$
\label{eq: discrete linear data matrix}
\begin{aligned}
    X'&=
    \begin{bmatrix}
        x_1' & x_2' & \dots & x_N'
    \end{bmatrix}, \\
    X&=
    \begin{bmatrix}
        x_1 & x_2 & \dots & x_N
    \end{bmatrix}, \\
    U&=
    \begin{bmatrix}
        u_1 & u_2 & \dots & u_N
    \end{bmatrix}.
\end{aligned}
$$
The policy and Lyapunov function are the same as those in the continuous-time case, which are in the form of \eqref{eq: linear policy} and \eqref{eq: quadratic Lyapunov}, respectively. To verify the condition in Theorem \ref{Lyapunov's second method for stability discrete}, we first compute the time difference of Lyapunov function:
\begin{equation}
\label{eq: Lyapunov difference}
    V(x')-V(x)=x^\top((A+BK)^\top P(A+BK)-P)x.
\end{equation}
To verify whether $V(x')-V(x)<0$ holds is equivalent to check
\begin{equation}
\label{eq: Lyapunov difference matrix}
    Q\triangleq(A+BK)^\top P(A+BK)-P\prec0.
\end{equation}
To this end, we need to replace matrices $A$ and $B$ with data. Similar to the case of continuous-time linear systems, we require that the matrix $Z$ defined in \eqref{eq: Z matrix} has full row rank so that it has a right inverse $Z^+$. Therefore, we have
\begin{equation}
\label{eq: discrete linear system representation}
\begin{bmatrix}
    A & B
\end{bmatrix}
=X'Z^+,
\end{equation}
The dissipation matrix becomes
\begin{equation}
\label{eq: discrete linear stability criterion}
    Q=\left(X'Z^+
    \begin{bmatrix}
        I \\ K
    \end{bmatrix}
    \right)^\top P
    X'Z^+
    \begin{bmatrix}
        I \\ K
    \end{bmatrix}
    -P.
\end{equation}
If $Q$ is negative definite, the discrete-time system is stable.

\subsection{Difference and connection with existing methods}
This section reviews a few representative works on the datatic stability of linear systems. Their problem formulations are slightly different from ours. We divide them into two categories: stability verification of autonomous systems and stabilization of closed-loop systems~\cite{markovsky2021behavioral}. The former considers systems without control input or the case where stability verification and data collection use the same policy. The latter focuses on how to synthesize a control policy that renders the closed-loop system stable, whose intermediate results are helpful to us in understanding stability verification for linear systems.
% √TODO: 最后加一句，他们的中间结果对我们理解stability是有帮助的。

\subsubsection{Stability verification of autonomous systems}
Waarde et al.~\cite{van2020data} study the informative conditions for datatic stability of an autonomous system. They conclude that the necessary and sufficient condition for stability verification is that data is informative for system identification. A discrete-time linear autonomous system is considered:
\begin{equation}
    x'=Ax.
\end{equation}
Data can be written in the form of $X'=AX$, where $X'$ and $X$ are given by \eqref{eq: discrete linear data matrix}. They require that the matrix $X$ has full row rank, i.e., $\mathrm{rank}(X)=n$.
Then, it has a right inverse $X^+$, such that $XX^+=I$. Therefore, we can represent the system matrix $A$ with data:
\begin{equation}
\label{eq: autonomous system representation}
    A=X'X^+.
\end{equation}
Waarde et al. use characteristic root method to verify its stability: if the absolute values of all eigenvalues of matrix $A$ are less than one, the autonomous system is stable.

Its difference from our method is obvious: Waarde et al. consider autonomous systems while we consider closed-loop control systems. This allows our method to deal with the case where the policy for data collection and that for stability verification are not the same. In linear cases, there is a similarity between these two methods: both need to compute the right inverse of a data matrix to represent plant dynamics. The computation of the right inverse requires that the data matrix has full row rank, i.e., there is a sufficient number of linearly independent data points.

\subsubsection{Stabilization of closed-loop systems}
Persis et al.~\cite{de2019formulas} design a state feedback policy to stabilize a linear datatic control system. Although stabilization is different from stability verification, we find some connections between their analysis and our simplified method for linear systems. Persis et al. consider a discrete-time linear control system and use Lyapunov criterion to verify stability. The plant dynamics, data, policy, and Lyapunov function are in the form of \eqref{eq: discrete linear dynamics}, \eqref{eq: discrete linear data}, \eqref{eq: linear policy}, and \eqref{eq: quadratic Lyapunov}, respectively. They also require that the data matrix has full row rank, which is expressed as \eqref{eq: full row rank}. Under this condition, the plant dynamics can be represented by data in the same way as \eqref{eq: discrete linear system representation}. If the policy and the Lyapunov function are given, i.e., matrices $K$ and $P$ are given, stability can be verified by checking whether \eqref{eq: discrete linear stability criterion} holds. These results are the same as our method in linear systems.

Persis et al.~\cite{de2019formulas} take a step forward to study how to stabilize a linear system. Their aim is to find a matrix $K$ so that the closed-loop system is stable. Starting from \eqref{eq: discrete linear stability criterion} and using variable transformation, they cast the stabilization problem into a linear matrix inequality (LMI):
\begin{equation}
\label{eq: stability LMI}
    \begin{bmatrix}
        XQ & X'Q \\
        Q^\top X'^\top & XQ
    \end{bmatrix}
    \succ0.
\end{equation}
They prove that for any matrix $Q$ satisfying \eqref{eq: stability LMI}, the feedback gain $K=UQ(XQ)^{-1}$ must stabilize the linear system. Moreover, the Lyapunov function can be computed by $P=XQ$.
This result provides an effective method for choosing Lyapunov functions for linear datatic control systems.

\section{Practical Implementations}
The case of linear systems assumes a strong condition: the plant dynamics are known to be linear. In practice, however, even if the system is linear, we may not know it. Not to mention that for nonlinear systems, this condition itself does not hold. So in this section, we return to general  systems to discuss our stability criterion. Below we will discuss two problems: one is how to estimate the Lipschitz constant of a datatic system, and the other is how to solve QCLP more efficiently.

\subsection{Lipschitz constant estimation}
Assumption \ref{asmp:Lipschitz continuity} provides the basis for how to extend plant representation from known data points to unknown regions. However, strict extension under this assumption faces two challenges: (1) The Lipschitz constants $L_x$ and $L_u$ cannot be analytically computed from $f$ since $f$ is unknown. (2) The Lipschitz constants in Assumption \ref{asmp:Lipschitz continuity} are global, i.e., they are equal on any point. In practice, we find that global Lipschitz constants are often too loose to restrict the value of $f$. Specifically, the intersection of balls given by \eqref{eq: intersection of balls} is often very large so that the solution of \eqref{equ:QCLP} may easily exceed zero, failing stability verification. 
To solve these problems, we first use data to estimate Lipschitz constants, and then replace global Lipschitz constants with local ones. First, let us define local Lipschitz continuity.

\begin{assumption}[Local Lipschitz continuity]
\label{asmp: local Lipschitz continuity}
For all $x\in\calX$ and $u\in\calU$, there exists a positive real constant $\delta$, such that $f(x,u)$ restricted to $\mathcal{B}((x_0,u_0),\delta)$ is Lipschitz continuous with respect to $x$ and $u$, i.e., there exists positive real constants $L_x$ and $L_u$, such that
\begin{equation}
\label{eq: local Lipschitz continuity}
\begin{aligned}
    \forall (x_1,u&), \sbr{x_2,u}\in\mathcal{B}((x_0,u_0),\delta), \\
    &d\sbr{f\sbr{x_1,u}, f\sbr{x_2,u}} \le L_x d\sbr{x_1, x_2}, \\
    \forall (x,u_1&), \sbr{x,u_2}\in\mathcal{B}((x_0,u_0),\delta), \\
    &d\sbr{f\sbr{x,u_1}, f\sbr{x,u_2}} \le L_u d\sbr{u_1,u_2},
\end{aligned}
\end{equation}
where
$$\mathcal{B}((x_0,u_0),\delta)=\{(x,u)|d((x,u),(x_0,u_0))\leq\delta\}.$$
The distance between state-action pairs is computed by first concatenating state and action into a single vector and then computing the distance between the concatenated vectors.
\end{assumption}

From Assumption \ref{asmp: local Lipschitz continuity}, we know that the Lipschitz constants $L_x$ and $L_u$ are related to $x_0$ and $u_0$. Therefore, we must estimate the Lipschitz constants on every data point to compute the radius of its corresponding closed balls. Although this will increase computational complexity, it can effectively improve the calculation accuracy of local Lipschitz constants. The estimation of $L_x$ and $L_u$ leverages the fact that they are the smallest numbers that let \eqref{eq: local Lipschitz continuity} hold. For each data point $(x_i,u_i)$, we solve the following quadratic program (QP) to find its corresponding local Lipschitz constants:
\begin{equation}
\label{eq: Lipschitz estimation}
\begin{aligned}
    \min_{L_x,L_u} \quad& \lambda L_x^2+L_u^2 \\
    \st \quad& d(\dot{x}_i,\dot{x}_j)\le L_xd(x_i,x_j)+L_ud(u_i,u_j), \\
    & \forall j\in\mathcal{I}_i,
\end{aligned}
\end{equation}
where $\lambda>0$ is an adjustable hyper-parameter and $\mathcal{I}_i$ is the index set of neighboring data points:
$$\mathcal{I}_i=\{j|d((x_i,u_i),(x_j,u_j))\le\delta,1\le j\le N\}.$$
We denote the local Lipschitz constants obtained by solving \eqref{eq: Lipschitz estimation} at the $i$-th data point as $L_{x_i}$ and $L_{u_i}$. Problem \eqref{eq: Lipschitz estimation} finds the smallest constants $L_x$ and $L_u$ that satisfy the local Lipschitz continuity condition. The constraints are constructed using data points within a distance of $\delta$ to $(x_i,u_i)$. In practice, we set $\delta$ as a constant for all $i=1,2,\dots,N$. 

\subsection{Algorithm simplification and complexity analysis}
Problem \eqref{equ:QCLP} uses global Lipschitz constants to construct constraints on $\dot{x}$. Here, we are going to replace them with local Lipschitz constants. However, for non-neighboring data points of $x_i$, local Lipschitz constants from solving \eqref{eq: Lipschitz estimation} do not apply. Therefore, we only use neighboring data points to construct constraints on $\dot{x}$, yielding
\begin{equation}
\label{eq: neighboring QCLP}
\begin{aligned}
    \max_{\dot{x}} \quad & \eta=\dv{V(x)}{x^\top}\bigg|_{x=x_i}\dot{x} \\
    \st \quad & (\dot{x}-\dot{x}_j)^\top(\dot{x}-\dot{x}_j)\le r_{ij}^2, \forall j\in\mathcal{I}_i^\pi,
\end{aligned}
\end{equation}
where
$$\mathcal{I}_i^\pi=\{j|d((x_i,\pi(x_i)),(x_j,u_j))\le\delta,1\le j\le N\}.$$
The radius in each constraint is computed using the local Lipschitz constants at the corresponding data point:
$$r_{ij}=L_{x_j}d(x_i,x_j)+L_{u_j}d(\pi(x_i),u_j).$$
The optimal value of \eqref{eq: neighboring QCLP} must be greater than or equal to that of \eqref{equ:QCLP} because some constraints are removed. Therefore, if the optimal value of \eqref{eq: neighboring QCLP} is less than zero, the optimal value of \eqref{equ:QCLP} must also be less than zero, and the system is stable.

The pseudocode of $\eta$-testing is shown in Algorithm \ref{alg: eta-tesing}. The algorithm consists of two parts: a) Lipschitz constant estimation and b) stability index calculation. 
Both parts need to traverse all data points. In Lipschitz constant estimation, the algorithm first finds the neighboring index set $\mathcal{I}_i$ and then solves the QP \eqref{eq: Lipschitz estimation}. The Lipschitz constants $L_{x_i}$ and $L_{u_i}$ at each data point are stored for computing the radius in \eqref{eq: neighboring QCLP}. In stability index calculation, the neighboring index set $\mathcal{I}_i^\pi$ is first computed and then QCLP \eqref{eq: neighboring QCLP} is solved to obtain the optimal value $\eta^*$. The optimal value is compared with zero to verify stability. If the optimal values at all data points are less than zero, the system is stable.

\begin{algorithm}
\caption{$\eta$-testing}
\label{alg: eta-tesing}
\KwIn{Dataset $\mathcal{D}$, policy $\pi$.}
\tcp{Lipschitz constant estimation}
\For{$i=1,2,\dots,N$}{
    Find neighboring index set $\mathcal{I}_i$. \\
    Compute $L_{x_i}$ and $L_{u_i}$ by solving QP \eqref{eq: Lipschitz estimation}. \\
}
\tcp{Stability index calculation}
\For{$i=1,2,\dots,N$}{
    Find neighboring index set $\mathcal{I}_i^\pi$. \\
    Compute $\eta^*$ by solving QCLP \eqref{eq: neighboring QCLP}. \\
    \If{$\eta^*\geq0$}{
        \Return{stability not verified}
    }
}
\Return{stability verified}
\end{algorithm}

We analyze the time complexity of Algorithm \ref{alg: eta-tesing} with respect to the number of data points $N$, state dimension $n$, and action dimension $m$. In Lipschitz constant estimation, the time complexity of traversing all data points is $\mathcal{O}(N)$. Finding the neighboring index set needs to traverse all data points and compute their distance from the current point, which has a time complexity of $\mathcal{O}(N(n+m))$. Problem \eqref{eq: Lipschitz estimation} is a convex QP with two variables. The number of constraints is dependent on $\delta$ and the density of data points in state and action spaces. In practice, we can choose $\delta$ according to $N$, $n$, and $m$ to approximately control the number of neighboring points. Here, we assume that the number of neighboring points is upper bounded by $M$. According to Ye and Tse~\cite{ye1989extension}, the time complexity of solving problem \eqref{eq: Lipschitz estimation} is $\mathcal{O}(M^2)$. Thus, the time complexity of Lipschitz constant estimation is $\mathcal{O}(NM^2+NM(n+m))$. In stability index calculation, the time complexity of traversing all data points is $\mathcal{O}(N)$ and that of finding the neighboring index set is $\mathcal{O}(N(n+m))$. Problem \eqref{eq: neighboring QCLP} can be formulated as a second-order cone programming with $n$ variables and $M$ constraints at most. According to Lobo et al.~\cite{lobo1998applications}, the time complexity of solving this problem is $\mathcal{O}(M^{3/2}n^3)$. Thus, the time complexity of stability index calculation is $\mathcal{O}(NM^{3/2}n^3+NM(n+m))$. 

To sum up, the overall $\eta$-testing algorithm has a time complexity of $\mathcal{O}(NM^2+NM^{3/2}n^3+NM(n+m))$. It is easy to see that the time complexity grows linearly with the number of data points $N$ and action dimension $m$, quadratically with the number of neighboring points $M$, and cubically with state dimension $n$.

\section{Experiments}
In this section, we test our stability criterion in three datatic control systems, including both linear and nonlinear ones. For each system, we verify the closed-loop dynamics under two policies, one for stability and one for instability. We also test the case of critical stability in one of them.

\subsection{Oscillator}
Consider a Van der Pol oscillator with control inputs, which is a highly nonlinear system:
\begin{equation}
\label{eq: oscillator dynamics}
    \ddot{y}=-y-\frac{1}{2}\dot{y}(1-y^2)+u,
\end{equation}
where $y$ is the position coordinate and $x=[y,\dot{y}]^\top$ is the state. The equilibrium point is $x_e=0$. We specify a bounded state space for data collection and stability verification:
$$\mathcal{X}=[-1,1]\times[-1,1].$$
Data collection consists of three steps. First, we uniformly sample $N=10000$ states in the state space. Then, a data-collecting policy is used to compute an action in each state. Finally, the time derivatives of states are computed using \eqref{eq: oscillator dynamics}. The data-collecting policy is obtained by adding noise to the policy for verification, which will be defined in each experiment below. The noise is sampled from a uniform distribution in the range $[-0.01,0.01]$. The method for data collection is the same in all following experiments and we will omit its explanation. The Lyapunov function is constructed in a quadratic form \eqref{eq: quadratic Lyapunov}.

\begin{figure}
    \subfloat[Data points]{
        \includegraphics[trim=10 0 10 0, clip, width=0.47\linewidth]{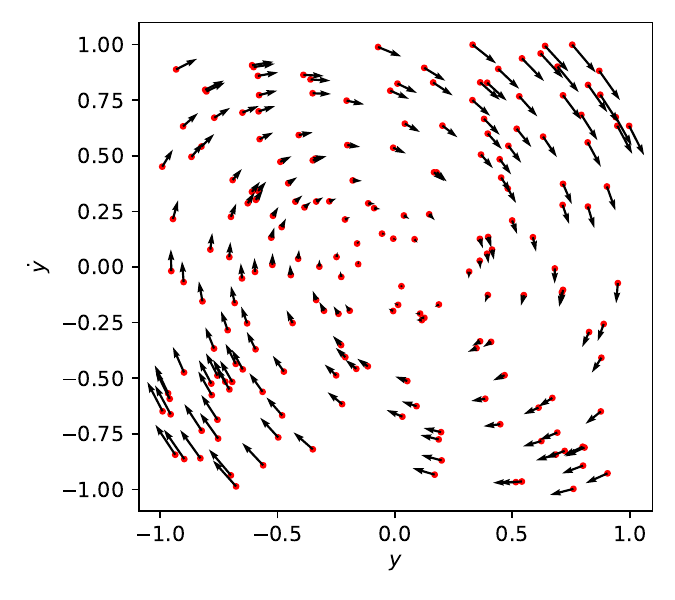}
    }
    \subfloat[Lyapunov function]{
        \includegraphics[trim=10 0 10 0, clip, width=0.53\linewidth]{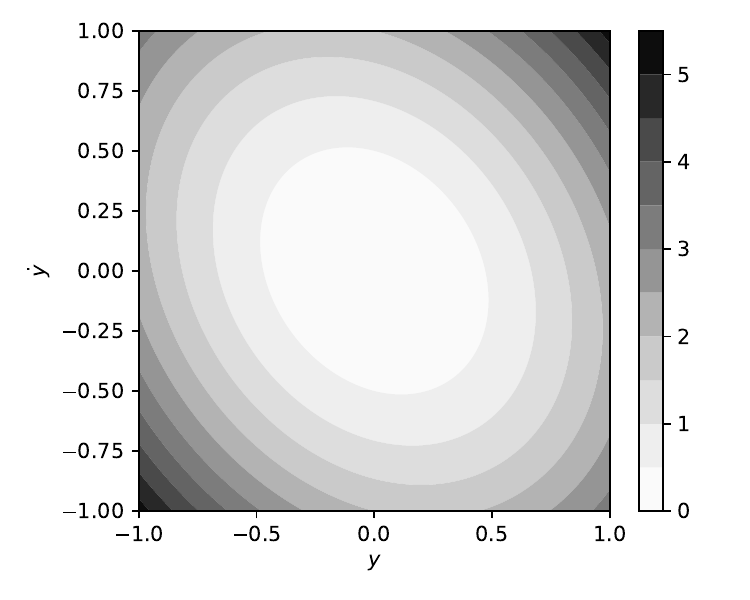}
    }
    \caption{Data points and Lyapunov function of oscillator for stability verification. (a) The red points represent states and the black arrows represent the corresponding time derivatives of states. The length of the arrows indicates the norm of time derivatives. (b) Contour map of Lyapunov function.}
    \label{fig: oscillator stable data and Lyapunov}
\end{figure}

For stability verification, the policy is chosen as
\begin{equation}
\label{eq: oscillator stable policy}
    u=-\frac{1}{2}y^2\dot{y}.
\end{equation}
The parameters of Lyapunov function are
$$P=
\begin{bmatrix}
    2.25 & 0.5 \\
    0.5 & 2
\end{bmatrix},$$
which are selected so that the stability conditions in Theorem \ref{thm: Lyapunov stability criterion} are satisfied under policy \eqref{eq: oscillator stable policy}. 
We visualize the collected data points (for visual clarity, we only choose $200$ points) and the Lyapunov function in the two-dimensional state space in Figure \ref{fig: oscillator stable data and Lyapunov}. It can be seen that the Lyapunov function is positive definite and takes the minimum value of zero at the equilibrium point.

\begin{figure}
    \subfloat[True value of $L_x$]{
        \includegraphics[trim=70 40 35 30, clip, width=0.5\linewidth]{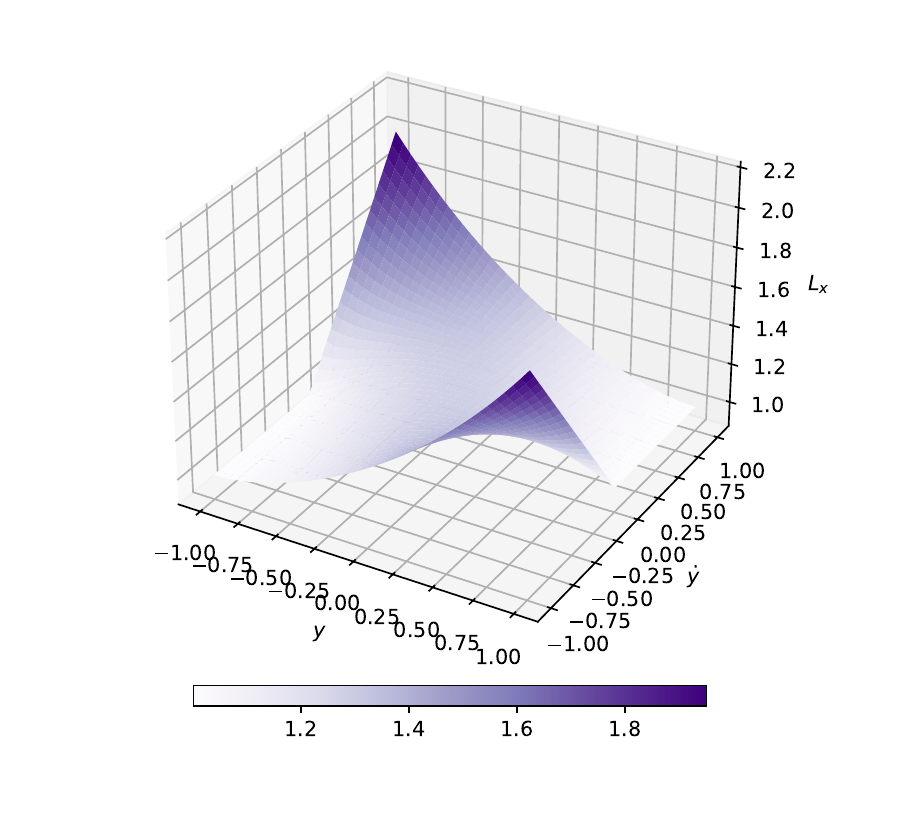}
    }
    \subfloat[Estimated value of $L_x$]{
        \includegraphics[trim=70 40 35 30, clip, width=0.5\linewidth]{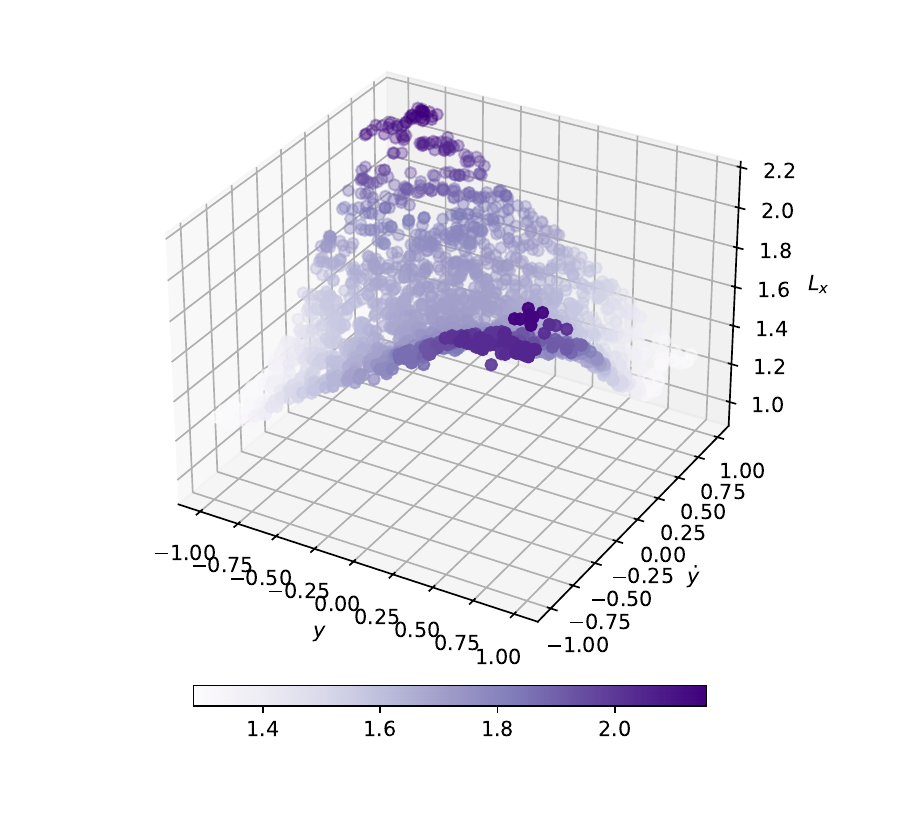}
    }\\
    \subfloat[True value of $L_u$]{
        \includegraphics[trim=70 40 35 30, clip, width=0.5\linewidth]{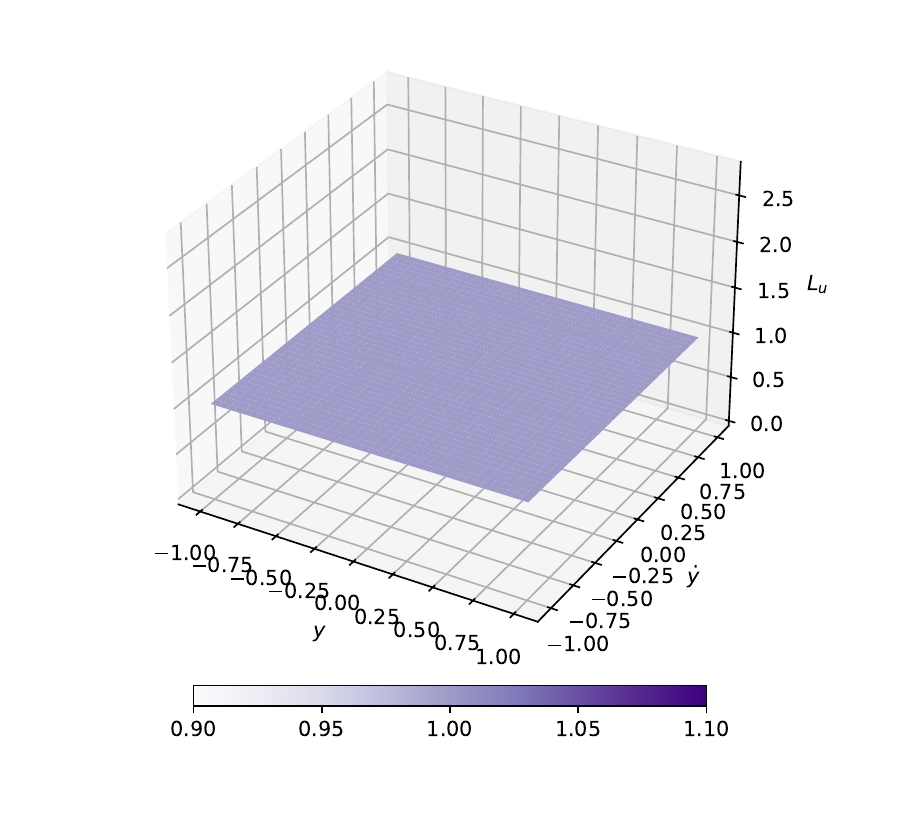}
    }
    \subfloat[Estimated value of $L_u$]{
        \includegraphics[trim=70 40 35 30, clip, width=0.5\linewidth]{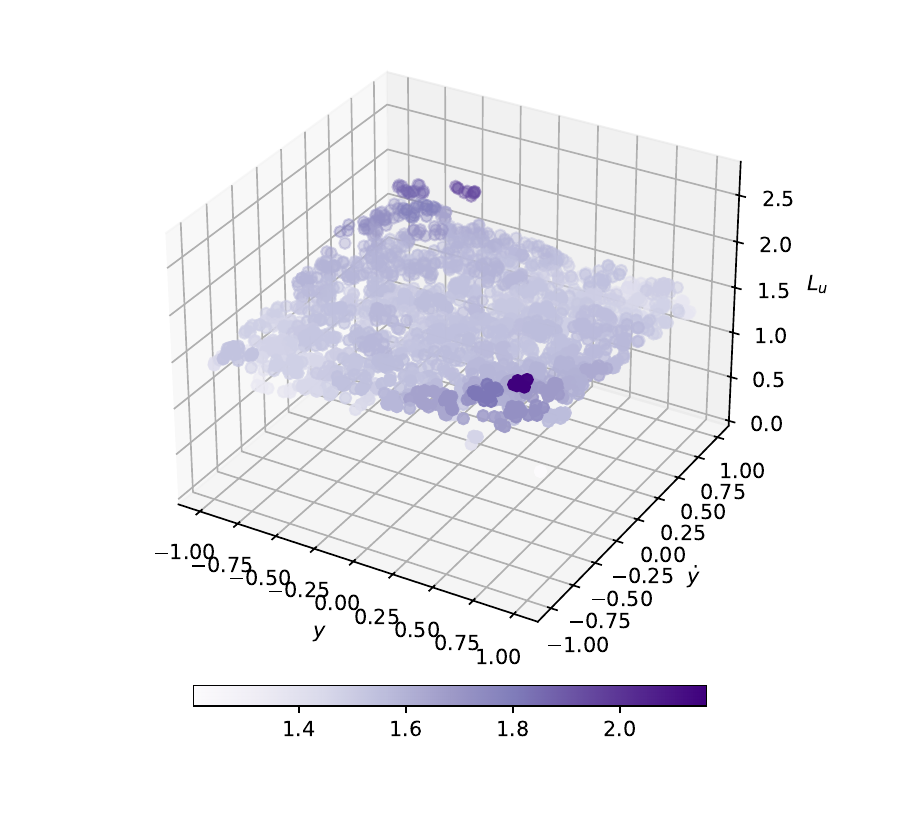}
    }
    \caption{True values and estimated values of Lipschitz constants of oscillator for stability verification. The estimated values are obtained by solving QP \eqref{eq: Lipschitz estimation}.}
    \label{fig: oscillator stable Lipschitz}
\end{figure}

To verify the closed-loop stability, we first estimate the Lipschitz constants using the first part of Algorithm \ref{alg: eta-tesing}. The size of neighboring region $\delta$ is set to $0.1$ and remains the same in all following experiments. We visualize the true and estimated Lipschitz constants in Figure \ref{fig: oscillator stable Lipschitz}. For $L_x$, the estimated value is consistent with the true value in terms of both numerical magnitude and changing trend. For $L_u$, the true value is a constant while the estimated value exhibits certain errors in some regions. Since the error magnitude is not too large compared to the true value, the performance of our stability verification method is not greatly affected.

\begin{figure}
    \subfloat[True value of $\dot{V}$]{
        \includegraphics[trim=70 40 35 30, clip, width=0.5\linewidth]{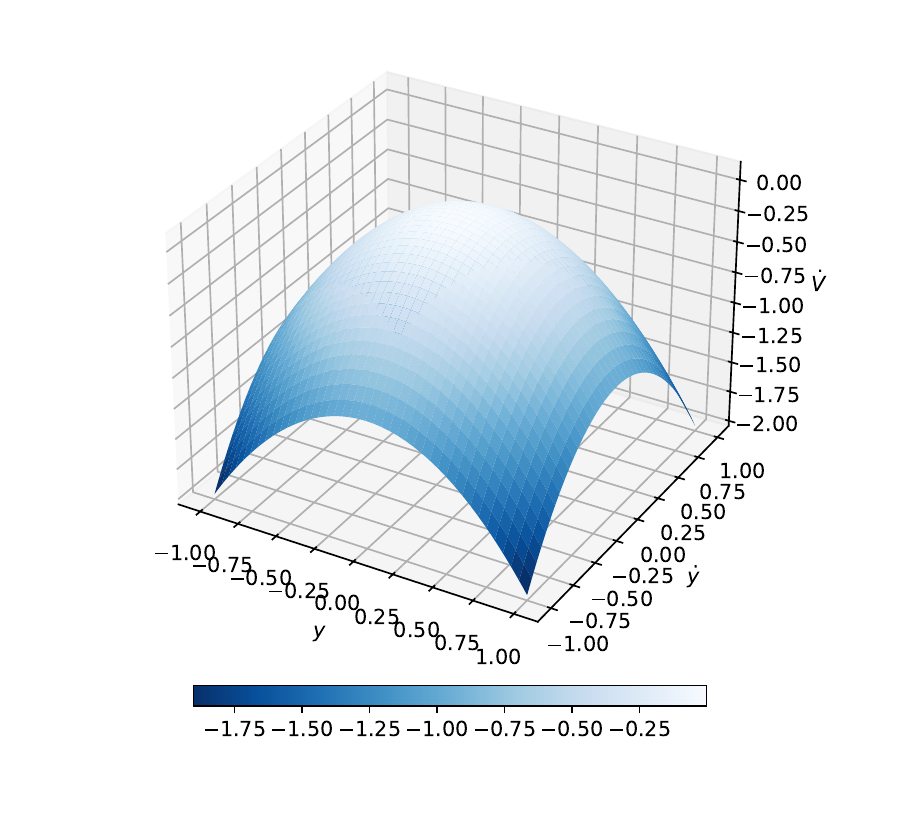}
    }
    \subfloat[Estimated value of $\dot{V}$]{
        \includegraphics[trim=70 40 35 30, clip, width=0.5\linewidth]{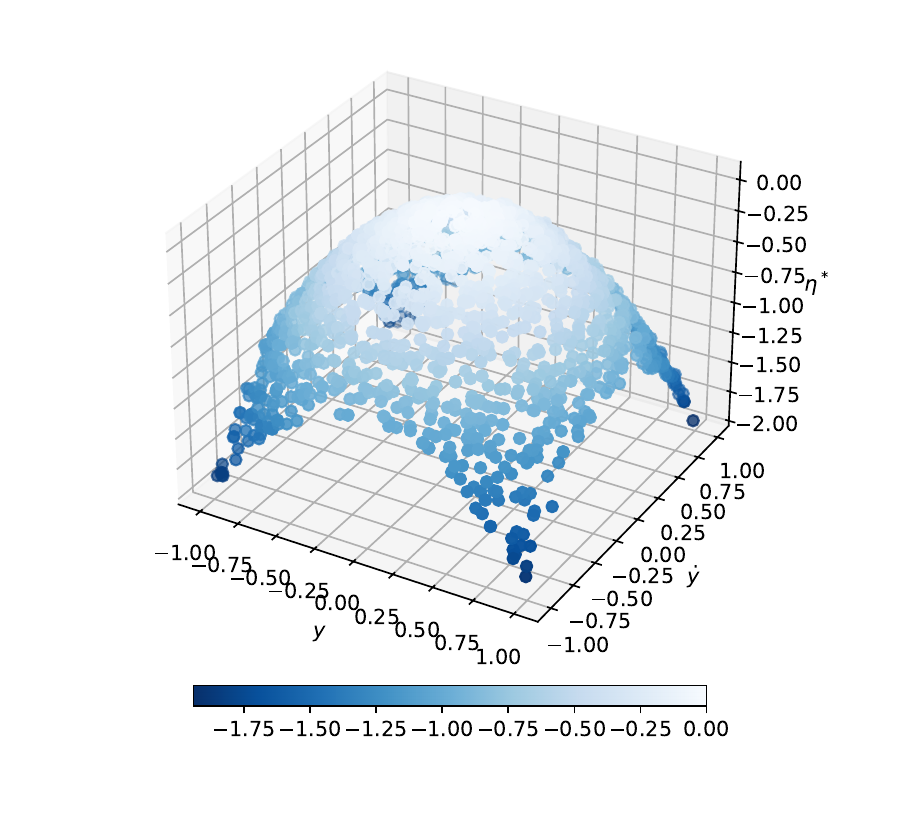}
    }
    \caption{True value and estimated value of time derivative of Lyapunov function of oscillator for stability verification. The estimated value is the optimal value $\eta^*$ of QCLP \eqref{equ:QCLP}.}
    \label{fig: oscillator stable v_dot}
\end{figure}

Then, we use the estimated Lipschitz constants to perform stability index calculation, which follows the second part of Algorithm \ref{alg: eta-tesing}. The true value and estimated value of time derivative of Lyapunov function are visualized in Figure \ref{fig: oscillator stable v_dot}. Figure \ref{fig: oscillator stable v_dot}(a) shows that the true time derivative is below zero in the entire state space, indicating that the underlying plant dynamics is stable. Figure \ref{fig: oscillator stable v_dot}(b) shows that the estimated time derivative $\eta^*$ obtained by solving QCLP \eqref{equ:QCLP} is also below zero in the entire state space, indicating that our $\eta$-testing algorithm successfully verifies the stability of the closed-loop system. It can be found through careful observation that the estimated time derivative is close but slightly larger than the true value. This is consistent with the fact that $\eta^*$ is a worst-case estimate in the intersection of closed balls of $\dot{x}$.

\begin{figure}
    \subfloat[Data points]{
        \includegraphics[trim=10 0 10 0, clip, width=0.47\linewidth]{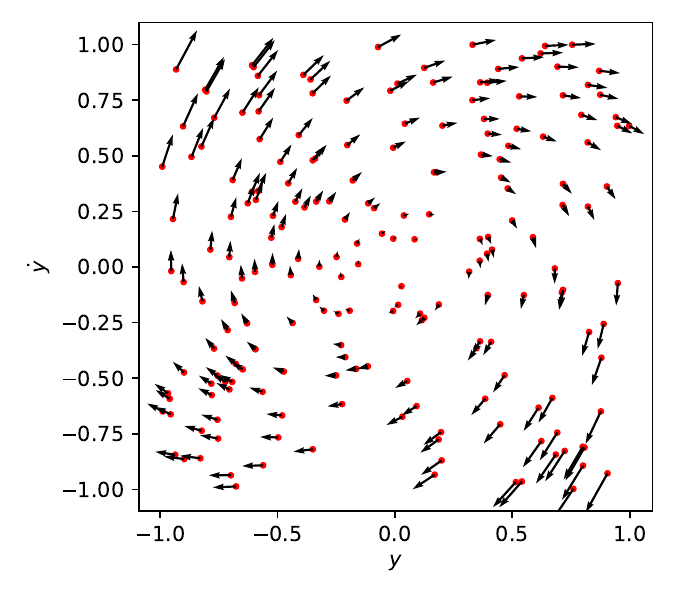}
    }
    \subfloat[Lyapunov function]{
        \includegraphics[trim=10 0 10 0, clip, width=0.53\linewidth]{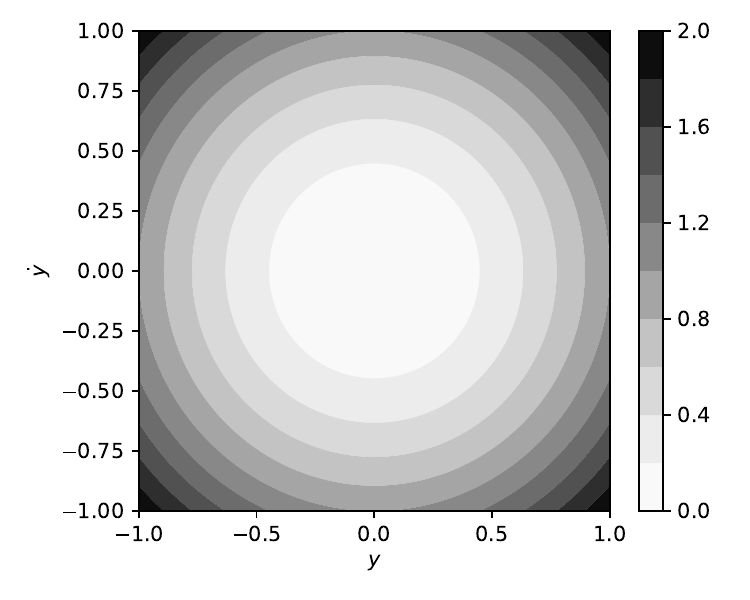}
    }
    \caption{Data points and Lyapunov function of oscillator for instability verification.}
    \label{fig: oscillator unstable data and Lyapunov}
\end{figure}

For instability verification, the policy is chosen as
\begin{equation}
\label{eq: oscillator unstable policy}
    u=\dot{y}.
\end{equation}
The parameters of Lyapunov function are set to
$$P=
\begin{bmatrix}
    1 & 0 \\
    0 & 1
\end{bmatrix},$$
which are selected so that the instability conditions in Theorem \ref{thm: Lyapunov instability criterion} are satisfied under \eqref{eq: oscillator unstable policy}.
We visualize collected data points and Lyapunov function in Figure \ref{fig: oscillator unstable data and Lyapunov}. It can be seen that the Lyapunov function is positive definite and takes the minimum value of zero at the equilibrium point.

\begin{figure}
    \subfloat[True value of $\dot{V}$]{
        \includegraphics[trim=70 40 35 30, clip, width=0.5\linewidth]{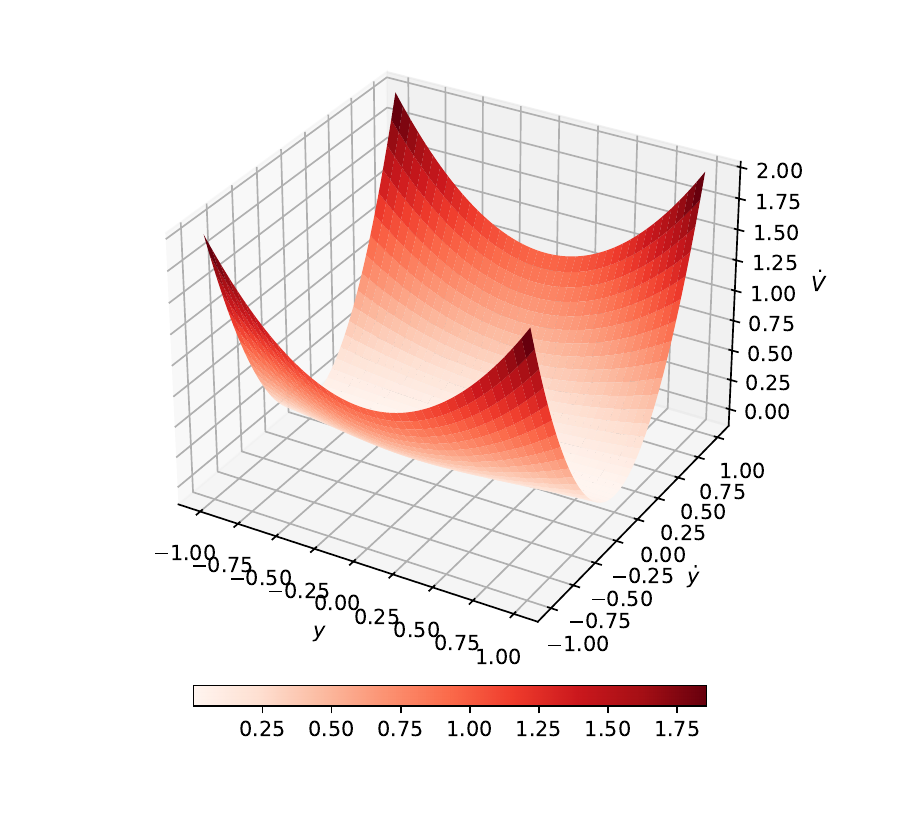}
    }
    \subfloat[Estimated value of $\dot{V}$]{
        \includegraphics[trim=70 40 35 30, clip, width=0.5\linewidth]{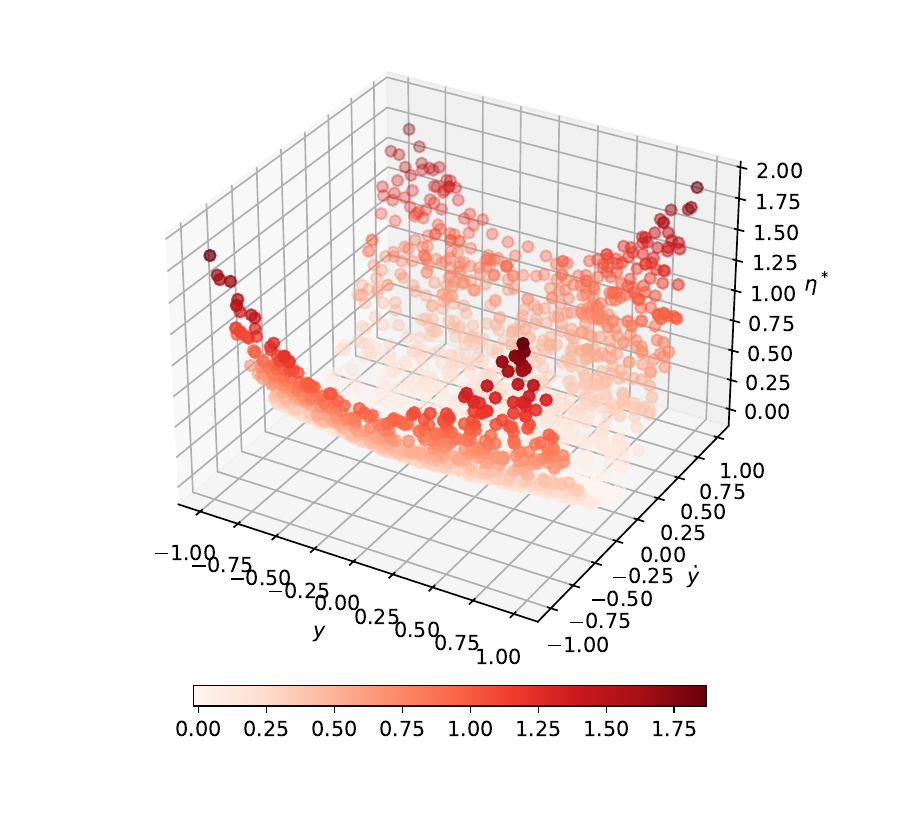}
    }
    \caption{True value and estimated value of time derivative of Lyapunov function of oscillator for instability verification. The estimated value is the optimal value $\eta^*$ of QCLP \eqref{eq: QCLP for instability}.}
    \label{fig: oscillator unstable v_dot}
\end{figure}

The estimation of Lipschitz constants is similar to that for stability verification and the results are omitted. We visualize the true value and estimated value of time derivative of Lyapunov function in Figure \ref{fig: oscillator unstable v_dot}. It shows that both the true and the estimated time derivatives are above zero in the entire state space, indicating that the closed-loop system is unstable and our $\eta$-testing algorithm successfully verifies the instability. As opposed to the case of stability verification, the estimated time derivative here is close but slightly smaller than the true value, which is due to the minimization in \eqref{eq: QCLP for instability} instead of maximization.

\subsection{2DOF vehicle model}
Consider a linear 2DOF vehicle model in the form of \eqref{eq: linear dynamics} with:
\begin{equation}
\begin{aligned}
    x&=
    \begin{bmatrix}
        y & \phi & v & \omega
    \end{bmatrix}^\top,
    u=\delta, \\
    A&=
    \begin{bmatrix}
        0 & u & 1 & 0 \\
        0 & 0 & 0 & 1 \\
        0 & 0 & \frac{k_f+k_r}{mu} & \frac{k_fl_f-k_rl_r}{mu}-u \\
        0 & 0 & \frac{k_fl_f-k_rl_r}{I_zu} & \frac{k_fl_f^2+k_rl_r^2}{I_zu}
    \end{bmatrix}, \\
    B&=
    \begin{bmatrix}
        0 & 0 & -\frac{k_f}{m} & -\frac{k_fl_f}{I_z}
    \end{bmatrix}^\top,
\end{aligned}
\end{equation}
where $y$ is the lateral position, $\phi$ is the heading angle, $v$ is the lateral velocity, $\omega$ is the raw rate, and $\delta$ is the front wheel angle. The equilibrium point is $x_e=0$. The vehicle parameters are listed in Table \ref{tab: vehicle parameters}. We specify a bounded state space for data collection and stability verification:
$$\mathcal{X}=[-1,1]\times\left[-\frac{\pi}{4},\frac{\pi}{4}\right]\times[-0.1,0.1]\times[-0.1,0.1].$$
The Lyapunov function is in a quadratic form \eqref{eq: quadratic Lyapunov} and the policy is a state feedback law \eqref{eq: linear policy}.

\begin{table}[htbp]
\begin{threeparttable}
    \centering
    \caption{Vehicle parameters.}
    \label{tab: vehicle parameters}
    \begin{tabular}{lll}
        \toprule
        Explanation & Symbol & Value \\
        \midrule
        Front wheel cornering stiffness & $k_f$ & $-80000 \ \mathrm{N/rad}$ \\
        Rear wheel cornering stiffness & $k_r$ & $-80000 \ \mathrm{N/rad}$ \\
        Distance from CG$^*$ to front axle & $l_f$ & $1.1 \ \mathrm{m}$ \\
        Distance from CG to rear axle & $l_f$ & $1.9 \ \mathrm{m}$ \\
        Mass & $m$ & $2000 \ \mathrm{kg}$ \\
        Polar moment of inertia at CG & $I_z$ & $2000 \ \mathrm{kg\cdot m^2}$ \\
        Longitudinal speed & $u$ & $5 \ \mathrm{m/s}$ \\
        \bottomrule
    \end{tabular}
    \begin{tablenotes}
        \item[*] CG stands for center of gravity.
    \end{tablenotes}
\end{threeparttable}
\end{table}

\begin{figure}
    \subfloat[Data points]{
        \includegraphics[trim=10 0 10 0, clip, width=0.47\linewidth]{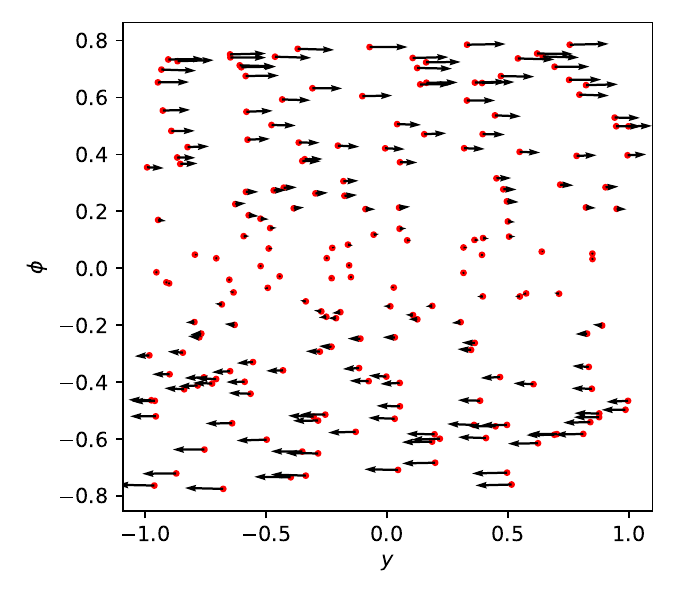}
    }
    \subfloat[Lyapunov function]{
        \includegraphics[trim=10 0 10 0, clip, width=0.53\linewidth]{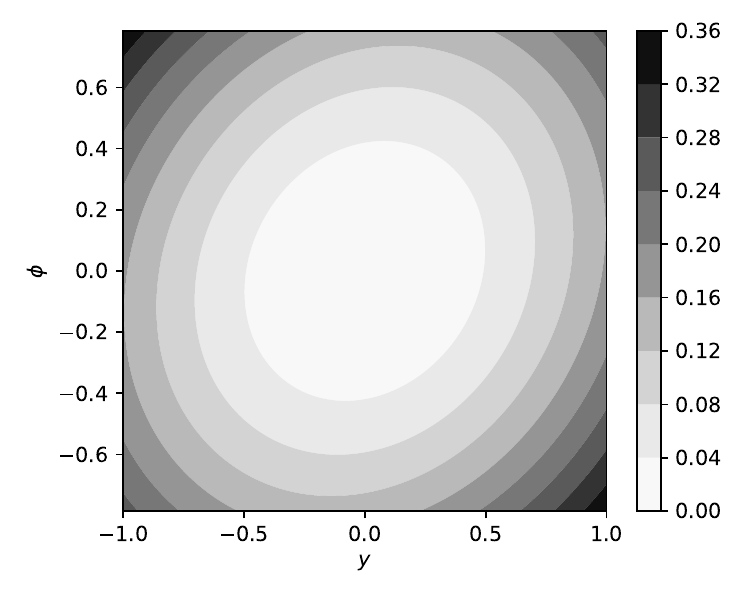}
    }
    \caption{Data points and Lyapunov function of vehicle for stability verification.}
    \label{fig: vehicle stable data and Lyapunov}
\end{figure}

For stability verification, the policy is chosen as a linear quadratic regulator, and the Lyapunov function is chosen as the value function of policy. The parameter matrix of Lyapunov function is obtained by solving an algebraic Riccati equation:
$$A^\top P+PA-PBR^{-1}B^\top P+Q=0,$$
where
$$Q=\mathrm{diag}([1,1,0.01,0.01]), \quad R=0.01.$$
The policy feedback gain is
$$K=-R^{-1}B^\top P.$$
We visualize collected data points and Lyapunov function in the first two dimensions of state space in Figure \ref{fig: vehicle stable data and Lyapunov}. It can be seen that the Lyapunov function is positive definite and takes the minimum value of zero at the equilibrium point.

\begin{figure}
    \subfloat[True value of $L_x$]{
        \includegraphics[trim=70 40 35 30, clip, width=0.5\linewidth]{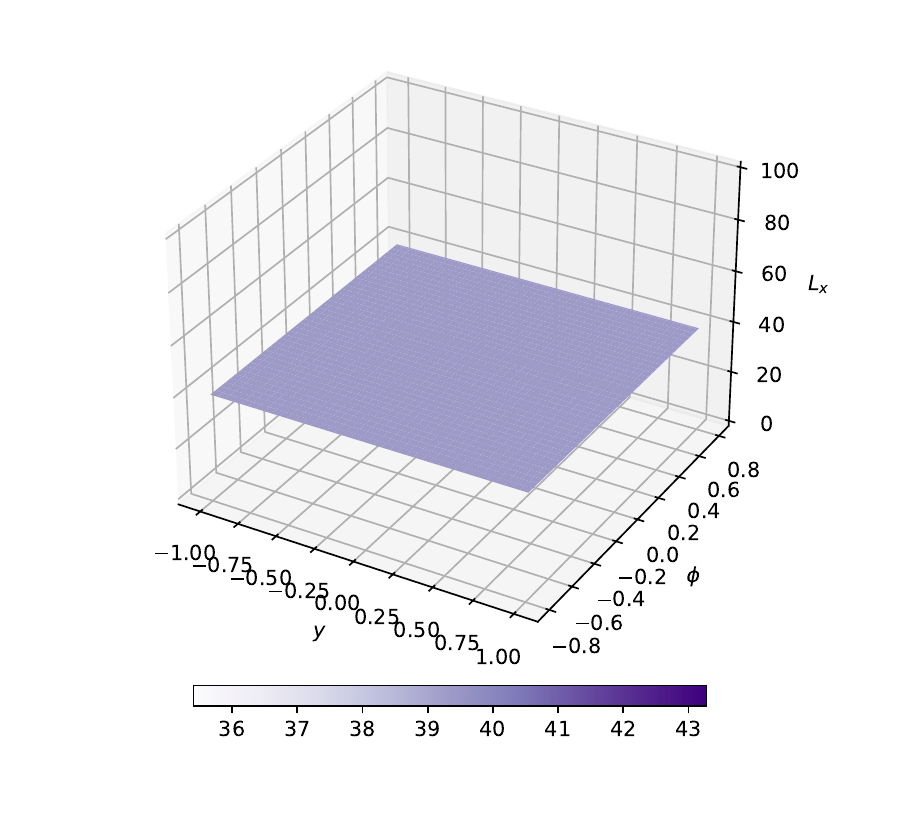}
    }
    \subfloat[Estimated value of $L_x$]{
        \includegraphics[trim=70 40 35 30, clip, width=0.5\linewidth]{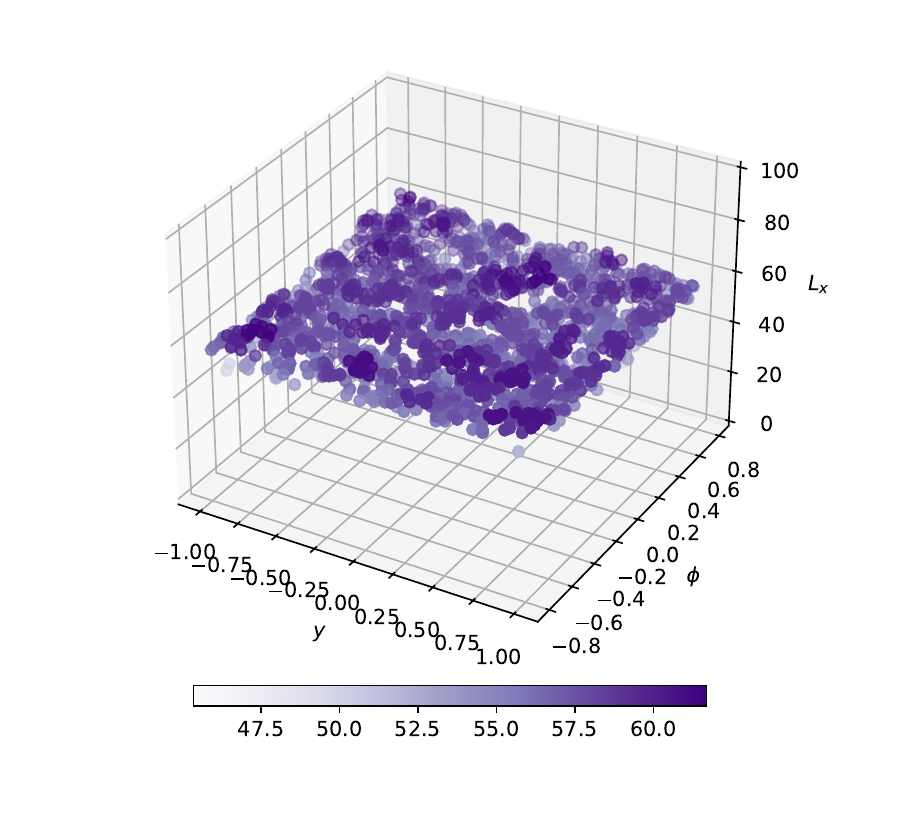}
    }\\
    \subfloat[True value of $L_u$]{
        \includegraphics[trim=70 40 35 30, clip, width=0.5\linewidth]{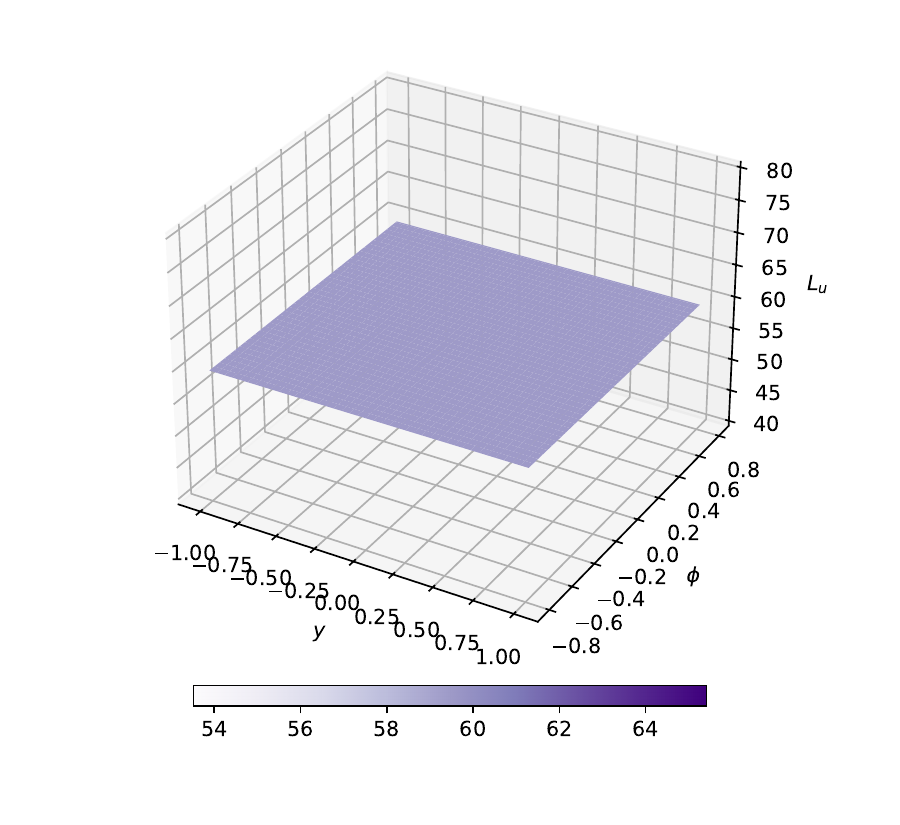}
    }
    \subfloat[Estimated value of $L_u$]{
        \includegraphics[trim=70 40 35 30, clip, width=0.5\linewidth]{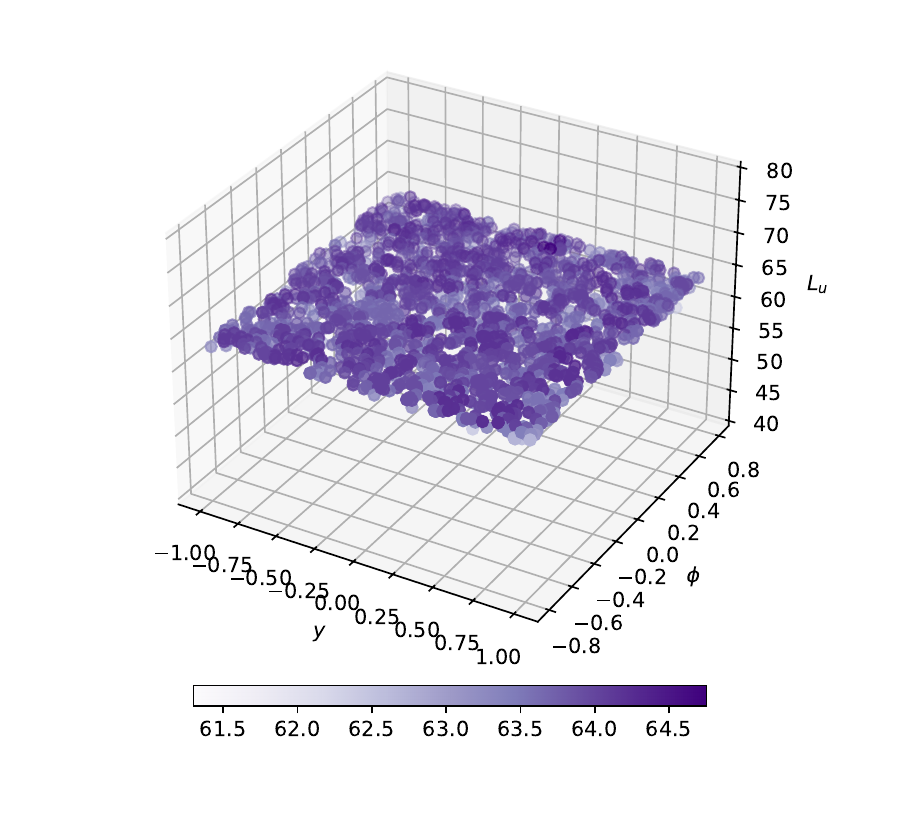}
    }
    \caption{True values and estimated values of Lipschitz constants of vehicle for stability verification.}
    \label{fig: vehicle stable Lipschitz}
\end{figure}

We first estimate the Lipschitz constants using the first part of Algorithm \ref{alg: eta-tesing} and visualize the results in Figure \ref{fig: vehicle stable Lipschitz}. It shows that the true Lipschitz constants of both state and action are constant throughout the state space, which is obvious since the system is linear. The estimated Lipschitz constants are slightly larger than the true values but can still be used for stability verification in this experiment.

\begin{figure}
    \subfloat[True value of $\dot{V}$]{
        \includegraphics[trim=70 40 35 30, clip, width=0.5\linewidth]{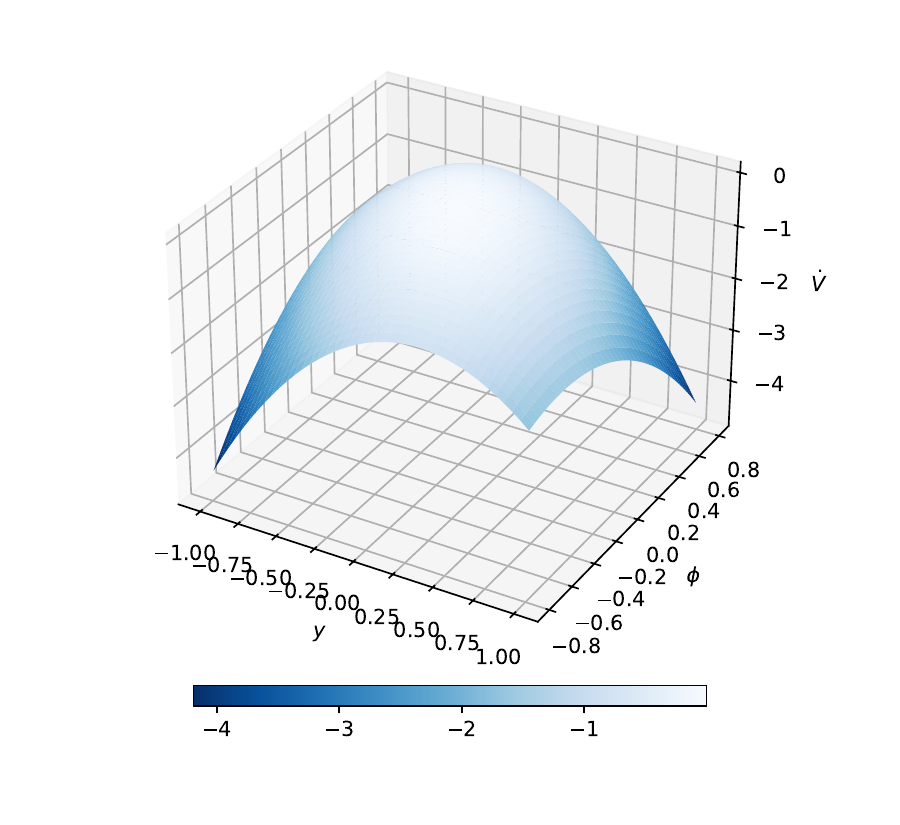}
    }
    \subfloat[Estimated value of $\dot{V}$]{
        \includegraphics[trim=70 40 35 30, clip, width=0.5\linewidth]{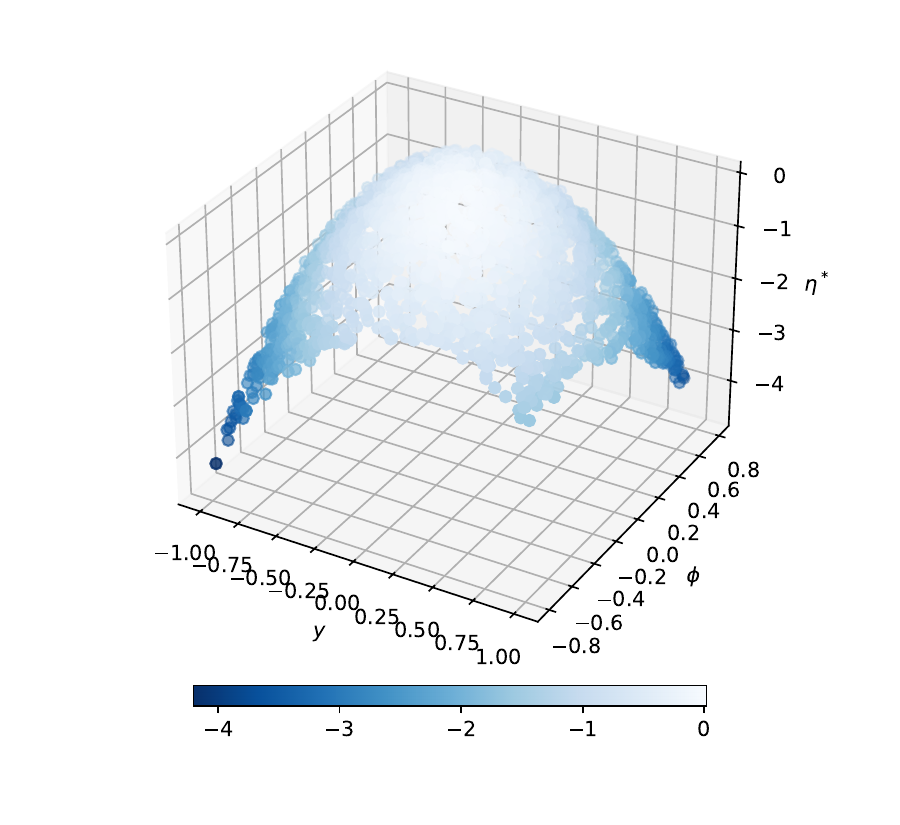}
    }
    \caption{True value and estimated value of time derivative of Lyapunov function of vehicle for stability verification.}
    \label{fig: vehicle stable v_dot}
\end{figure}

Then, we verify system stability using the second part of Algorithm \ref{alg: eta-tesing} and visualize the results in Figure \ref{fig: vehicle stable v_dot}. It shows that both the true and the estimated time derivatives of Lyapunov function are below zero in the entire state space, indicating that the closed-loop system is stable and our $\eta$-testing algorithm successfully verifies the stability.

\begin{figure}
    \subfloat[Data points]{
        \includegraphics[trim=10 0 10 0, clip, width=0.47\linewidth]{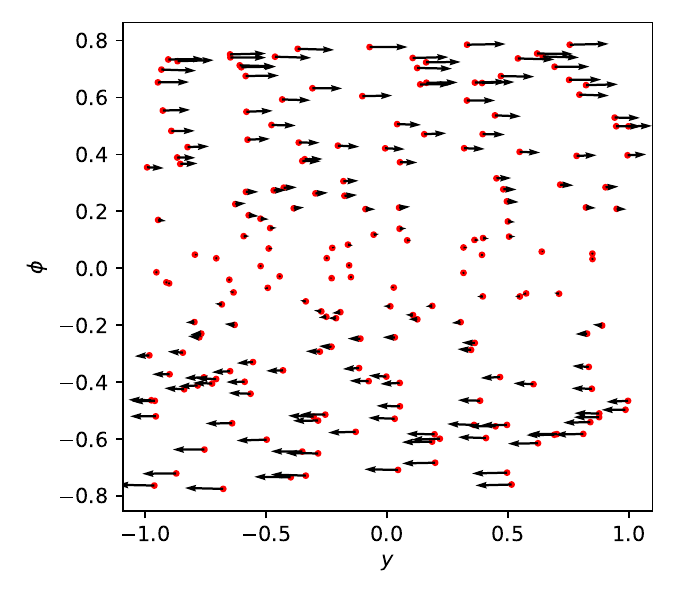}
    }
    \subfloat[Lyapunov function]{
        \includegraphics[trim=10 0 10 0, clip, width=0.53\linewidth]{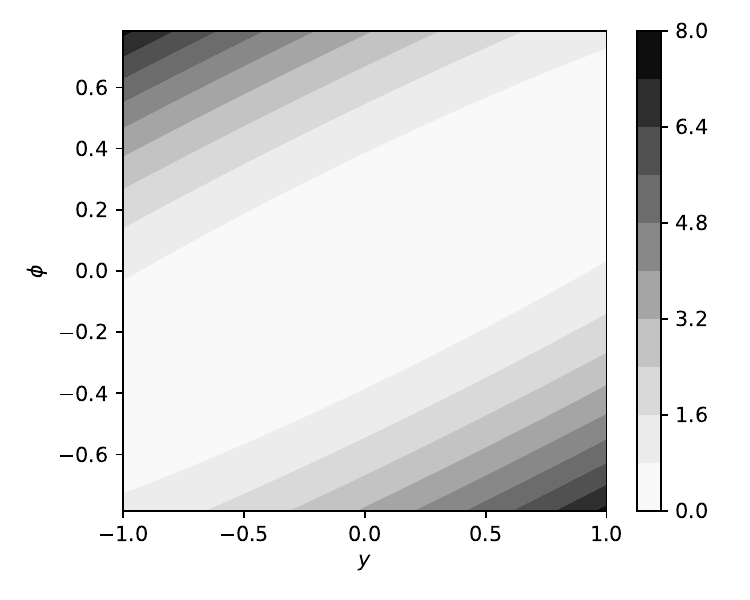}
    }
    \caption{Data points and Lyapunov function of vehicle for instability verification.}
    \label{fig: vehicle unstable data and Lyapunov}
\end{figure}

For instability verification, the Lyapunov function is also computed by solving the algebraic Riccati equation in stability verification but with the signs of $A$ and $B$ reversed. The policy feedback gain is computed with the sign of $B$ reversed. We visualize collected data points and Lyapunov function in Figure \ref{fig: vehicle unstable data and Lyapunov}. It can be seen that the Lyapunov function is positive definite and takes the minimum value of zero at the equilibrium point.

\begin{figure}
    \subfloat[True value of $\dot{V}$]{
        \includegraphics[trim=70 40 35 30, clip, width=0.5\linewidth]{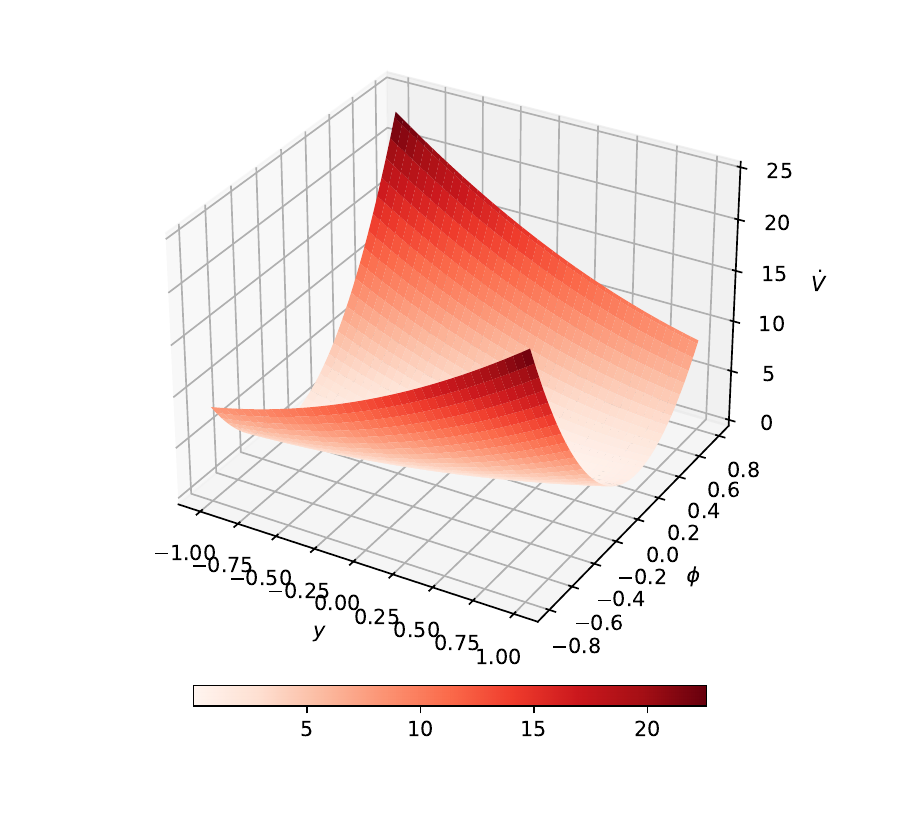}
    }
    \subfloat[Estimated value of $\dot{V}$]{
        \includegraphics[trim=70 40 35 30, clip, width=0.5\linewidth]{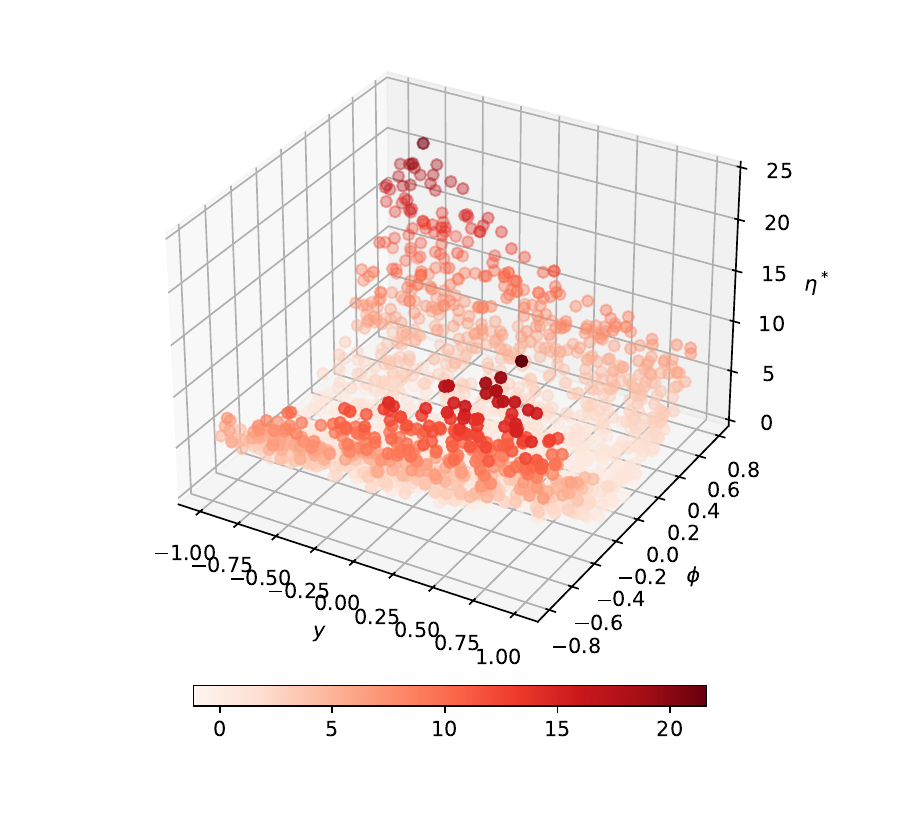}
    }
    \caption{True value and estimated value of time derivative of Lyapunov function of vehicle for instability verification.}
    \label{fig: vehicle unstable v_dot}
\end{figure}

The estimation of Lipschitz constants is similar to that for stability verification and the results are omitted. We visualize the true value and estimated value of time derivative of Lyapunov function in Figure \ref{fig: vehicle unstable v_dot}. It shows that both the true and the estimated time derivatives are above zero in the entire state space, indicating that the closed-loop system is unstable and our $\eta$-testing algorithm successfully verifies the instability.

\subsection{Pendulum}
Consider a pendulum with the following nonlinear dynamics:
\begin{equation}
    \ddot{\theta}=-\frac{3g}{2l}\sin\theta+\frac{3}{ml^2}u,
\end{equation}
where $\theta$ is the angle and $x=[\theta,\dot{\theta}]^\top$ is the state. The pendulum parameters are listed in Table \ref{tab: pendulum parameters}. For stability verification, the equilibrium point is $x_e=[0,0]^\top$, which is the lowest point of the pendulum, and the state space is
$$\mathcal{X}=\left[-\frac{\pi}{2},\frac{\pi}{2}\right]\times[-2,2].$$
For instability verification, the equilibrium point is $x_e=[\pi,0]^\top$, which is the highest point of the pendulum, and the state space is
$$\mathcal{X}=\left[\frac{\pi}{2},\frac{3\pi}{2}\right]\times[-2,2].$$
The Lyapunov function is constructed as follows:
\begin{equation}
\label{eq: pendulum Lyapunov}
    V(x)=\frac{1}{3}x^\top Px+\frac{g}{l}(1-\cos(\theta-\theta_e)),
\end{equation}
where $\theta_e$ is the angle of the equilibrium point and
$$P=\begin{bmatrix}
    p_{11} & p_{12} \\
    p_{12} & p_{22}
\end{bmatrix}$$
is a parameter matrix.

\begin{table}[htbp]
    \centering
    \caption{Pendulum parameters.}
    \label{tab: pendulum parameters}
    \begin{tabular}{lll}
        \toprule
        Explanation & Symbol & Value \\
        \midrule
        Mass & $m$ & $1 \ \mathrm{kg}$ \\
        Length & $l$ & $1 \ \mathrm{m}$ \\
        Gravitational acceleration & $g$ & $9.8 \ \mathrm{m/s^2}$ \\
        \bottomrule
    \end{tabular}
\end{table}

\begin{figure}
    \subfloat[Data points]{
        \includegraphics[trim=10 0 10 0, clip, width=0.47\linewidth]{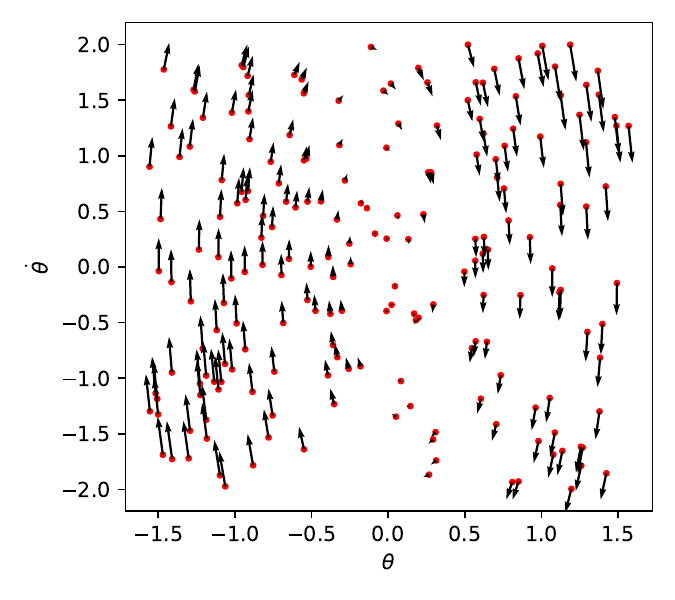}
    }
    \subfloat[Lyapunov function]{
        \includegraphics[trim=10 0 10 0, clip, width=0.53\linewidth]{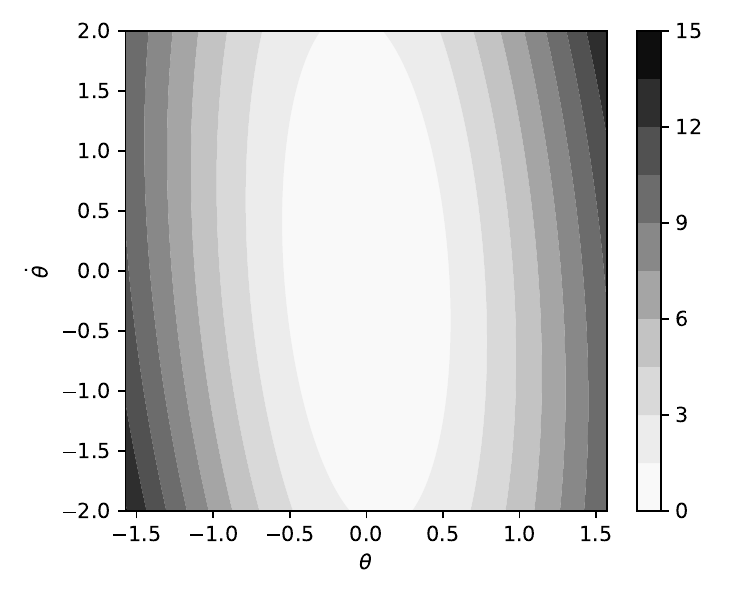}
    }
    \caption{Data points and Lyapunov function of pendulum for stability verification.}
    \label{fig: pendulum stable data and Lyapunov}
\end{figure}

For stability verification, the policy is an inverse proportional function of angular velocity:
\begin{equation}
\label{eq: pendulum stable policy}
    u=-k\dot{\theta},
\end{equation}
where $k=0.5$. The parameters of Lyapunov function are set to
$$p_{11}=\frac{9k^2}{2m^2l^4}, \quad p_{12}=\frac{3k}{2ml^2}, \quad p_{22}=1,$$
which are selected so that the stability conditions in Theorem \ref{thm: Lyapunov stability criterion} are satisfied under policy \eqref{eq: pendulum stable policy}. We visualize collected data points and Lyapunov function in Figure \ref{fig: pendulum stable data and Lyapunov}. It can be seen that the Lyapunov function is positive definite and takes the minimum value of zero at the equilibrium point.

\begin{figure}
    \subfloat[True value of $L_x$]{
        \includegraphics[trim=70 40 35 30, clip, width=0.5\linewidth]{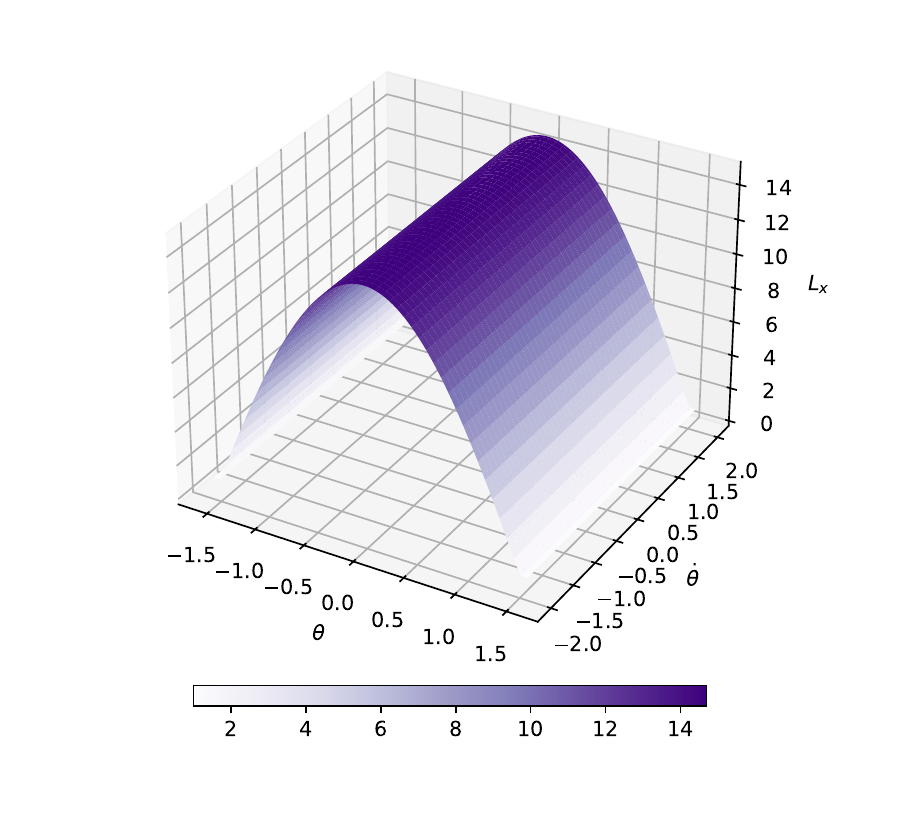}
    }
    \subfloat[Estimated value of $L_x$]{
        \includegraphics[trim=70 40 35 30, clip, width=0.5\linewidth]{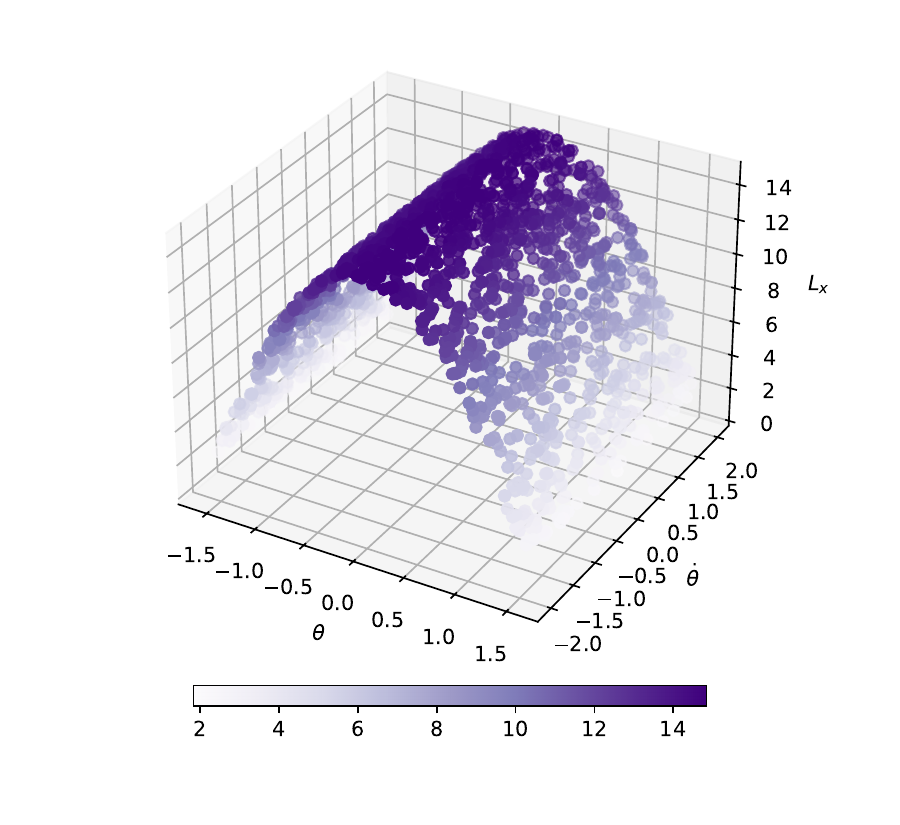}
    }\\
    \subfloat[True value of $L_u$]{
        \includegraphics[trim=70 40 35 30, clip, width=0.5\linewidth]{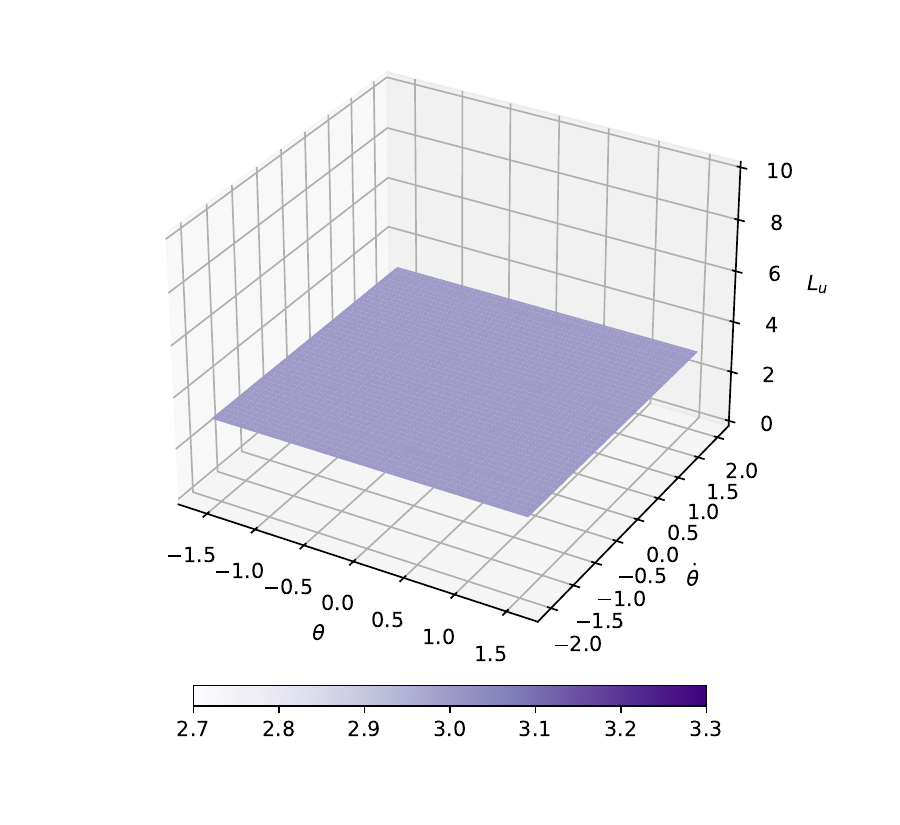}
    }
    \subfloat[Estimated value of $L_u$]{
        \includegraphics[trim=70 40 35 30, clip, width=0.5\linewidth]{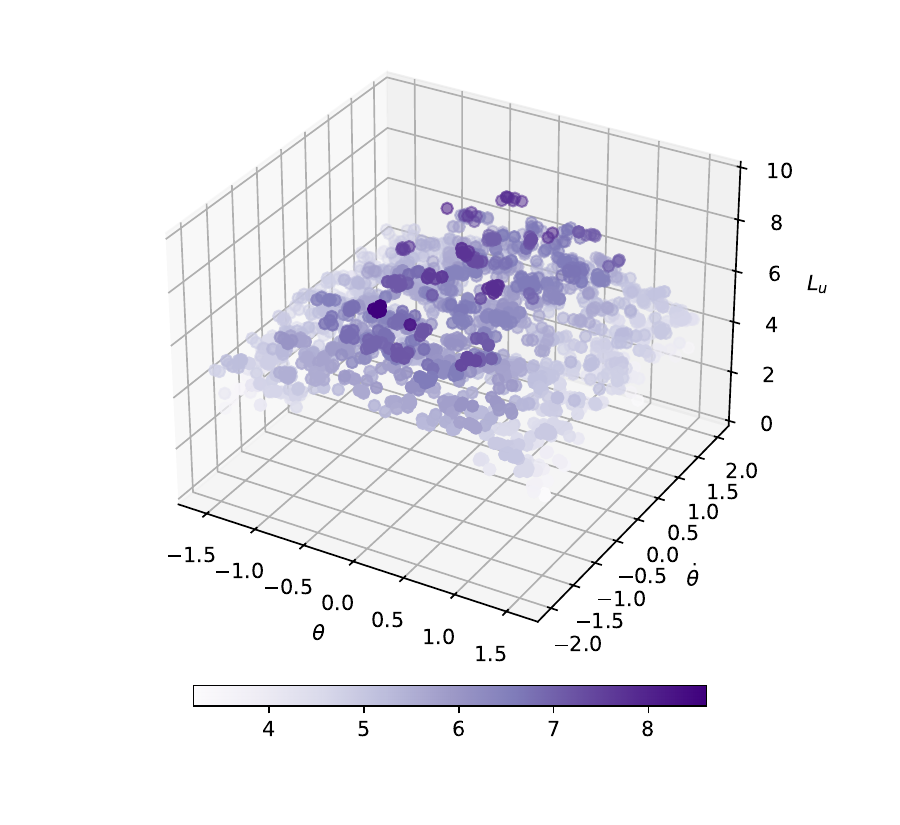}
    }
    \caption{True values and estimated values of Lipschitz constants of pendulum for stability verification.}
    \label{fig: pendulum stable Lipschitz}
\end{figure}

We first estimate the Lipschitz constants and visualize the results in Figure \ref{fig: pendulum stable Lipschitz}. Similar to the previous two experiments, the estimate of state-based Lipschitz constant is relatively accurate while that of action is slightly larger than the true value. 

\begin{figure}
    \subfloat[True value of $\dot{V}$]{
        \includegraphics[trim=70 40 35 30, clip, width=0.5\linewidth]{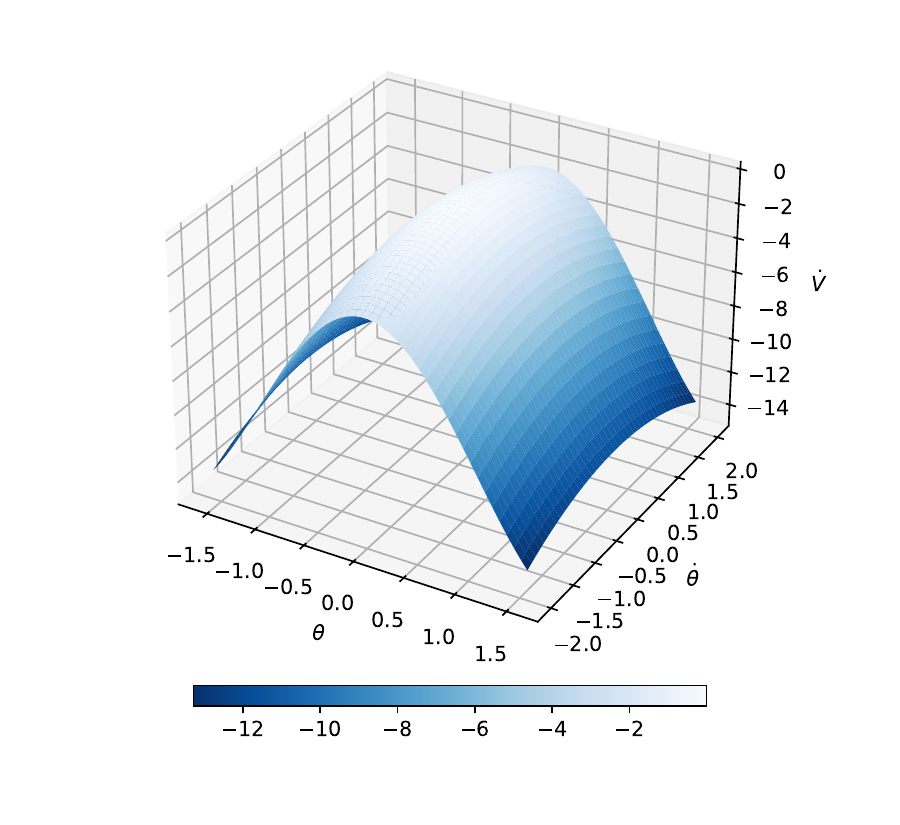}
    }
    \subfloat[Estimated value of $\dot{V}$]{
        \includegraphics[trim=70 40 35 30, clip, width=0.5\linewidth]{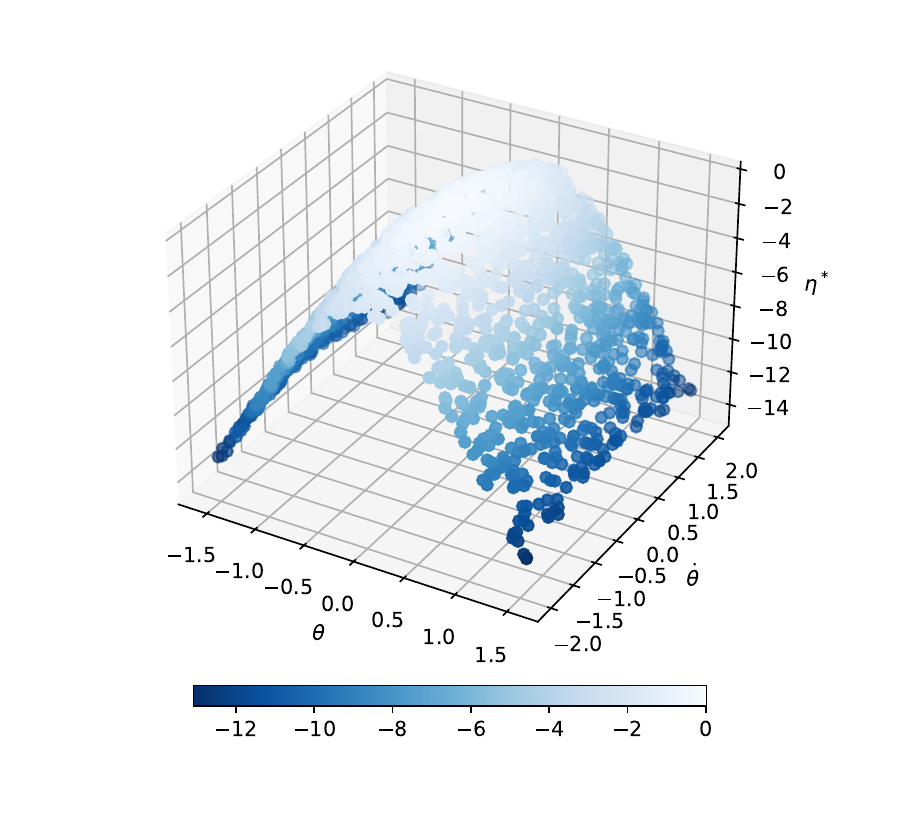}
    }
    \caption{True value and estimated value of time derivative of Lyapunov function of pendulum for stability verification.}
    \label{fig: pendulum stable v_dot}
\end{figure}

Then, we verify system stability and visualize the results in Figure \ref{fig: pendulum stable v_dot}. It shows that both the true and the estimated time derivatives of Lyapunov function are below zero in the entire state space, indicating that the closed-loop system is stable and our $\eta$-testing algorithm successfully verifies the stability.

\begin{figure}
    \subfloat[Data points]{
        \includegraphics[trim=10 0 10 0, clip, width=0.47\linewidth]{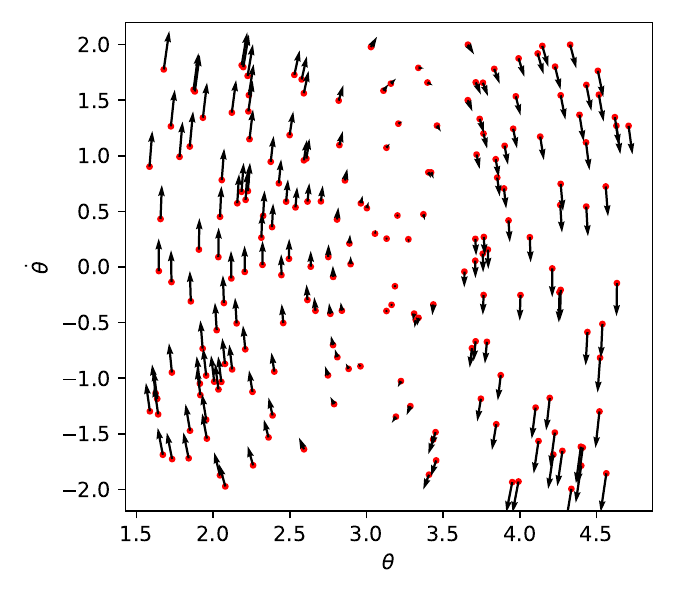}
    }
    \subfloat[Lyapunov function]{
        \includegraphics[trim=10 0 10 0, clip, width=0.53\linewidth]{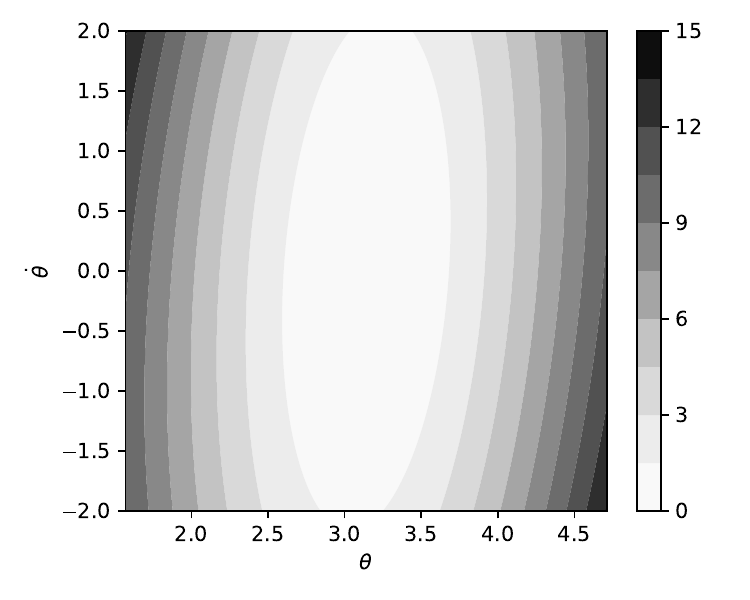}
    }
    \caption{Data points and Lyapunov function of vehicle for instability verification.}
    \label{fig: pendulum unstable data and Lyapunov}
\end{figure}

For instability verification, the policy is a proportional function of angular velocity plus a term for resisting gravity since the equilibrium point is on the top:
\begin{equation}
\label{eq: pendulum unstable policy}
    u=k\dot{\theta} + mgl\sin\theta.
\end{equation}
The Lyapunov function is still in the form of \eqref{eq: pendulum Lyapunov} and its parameters are
$$p_{11}=\frac{9k^2}{2m^2l^4}, \quad p_{12}=-\frac{3k}{2ml^2}, \quad p_{22}=1,$$
which are selected so that the instability conditions in Theorem \ref{thm: Lyapunov instability criterion} are satisfied under policy \eqref{eq: pendulum unstable policy}. We visualize collected data points and Lyapunov function in Figure \ref{fig: pendulum unstable data and Lyapunov}. It can be seen that the Lyapunov function is positive definite and takes the minimum value of zero at the equilibrium point.

\begin{figure}
    \subfloat[True value of $\dot{V}$]{
        \includegraphics[trim=70 40 35 30, clip, width=0.5\linewidth]{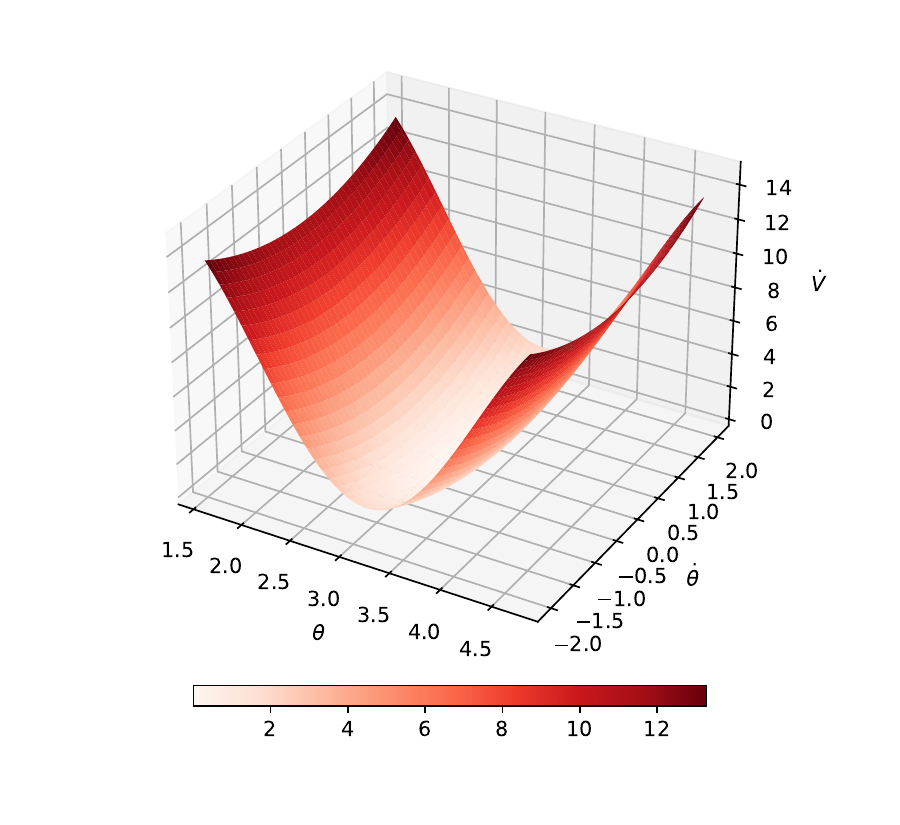}
    }
    \subfloat[Estimated value of $\dot{V}$]{
        \includegraphics[trim=70 40 35 30, clip, width=0.5\linewidth]{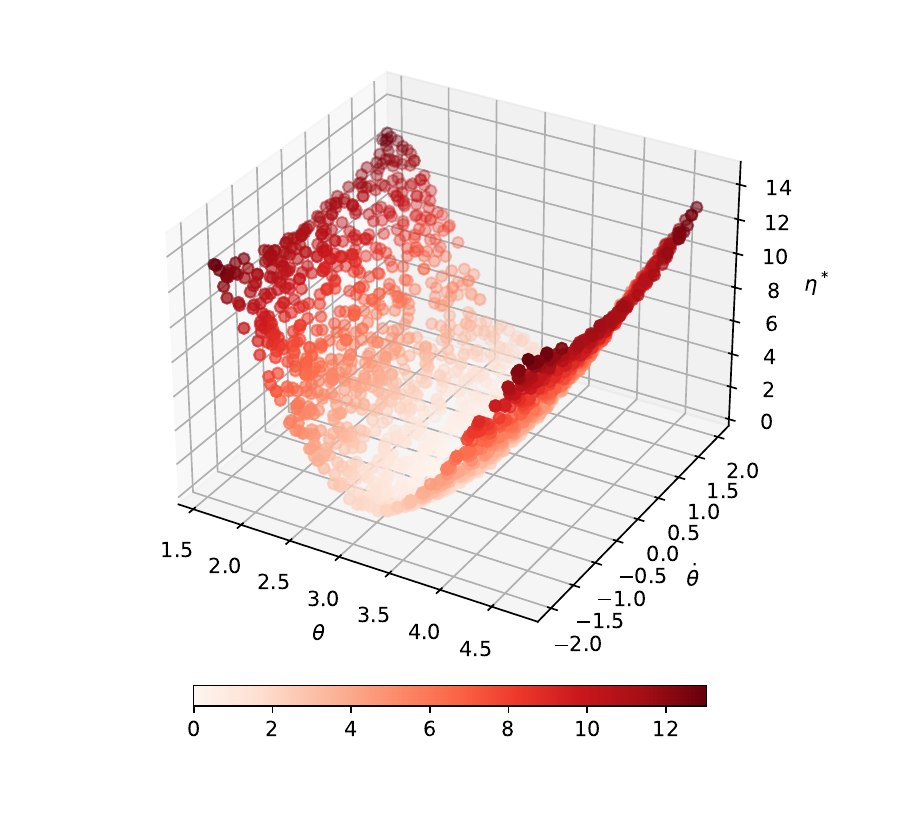}
    }
    \caption{True value and estimated value of time derivative of Lyapunov function of vehicle for instability verification.}
    \label{fig: pendulum unstable v_dot}
\end{figure}

The estimation of Lipschitz constants is similar to that for stability verification and the results are omitted. We visualize the true value and estimated value of time derivative of Lyapunov function in Figure \ref{fig: pendulum unstable v_dot}. It shows that both the true and the estimated time derivatives are above zero in the entire state space, indicating that the closed-loop system is unstable and our $\eta$-testing algorithm successfully verifies the instability.

\begin{figure}
    \subfloat[Stability verification]{
        \includegraphics[trim=70 40 35 30, clip, width=0.5\linewidth]{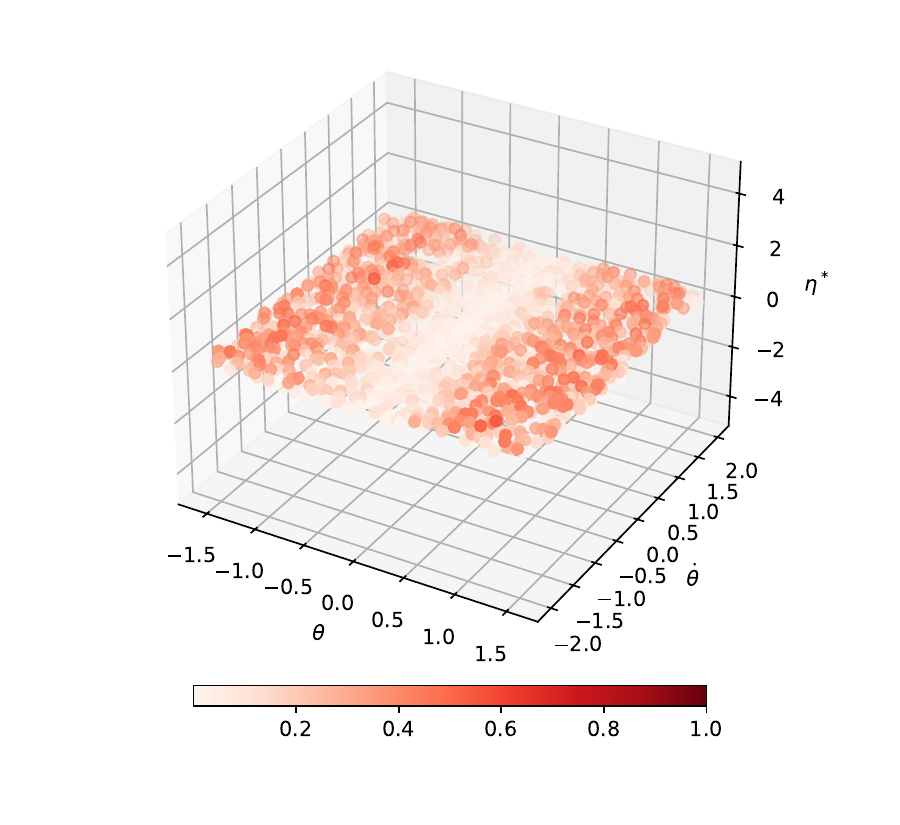}
    }
    \subfloat[Instability verification]{
        \includegraphics[trim=70 40 35 30, clip, width=0.5\linewidth]{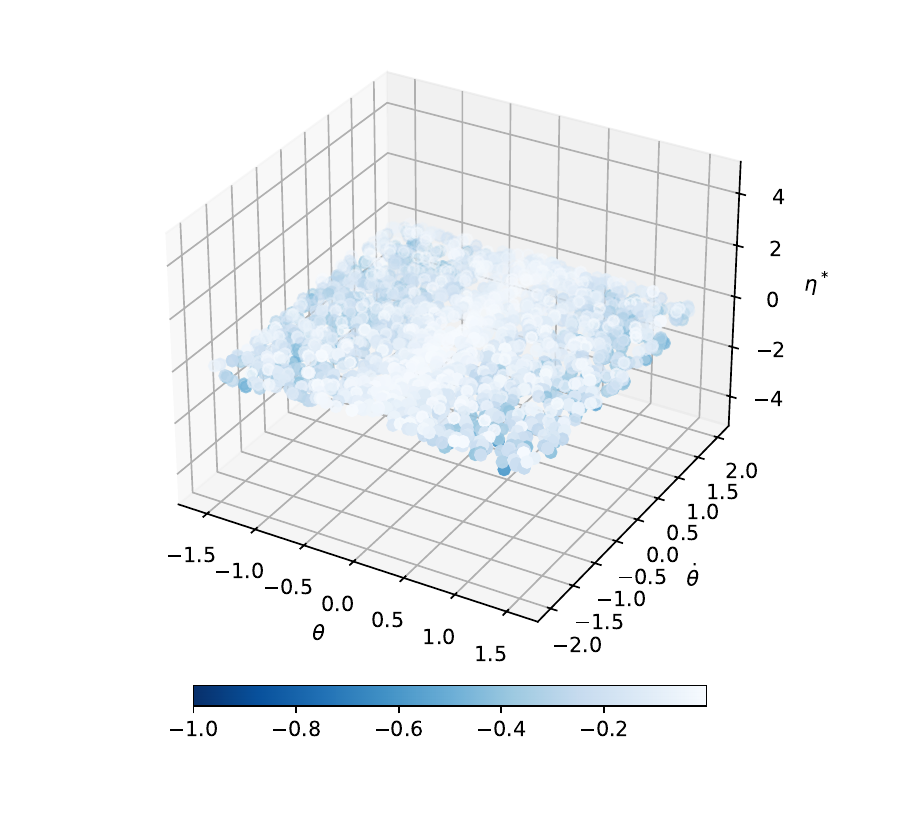}
    }
    \caption{Time derivative of Lyapunov function of vehicle estimated by stability and instability verification.}
    \label{fig: pendulum critical v_dot}
\end{figure}

In addition, we verify a case of critical stability in this system. We remove the control policy and let the pendulum swing freely, i.e., $u=0$. The Lyapunov function is the mechanical energy of the closed-loop system, which is also in the form of \eqref{eq: pendulum Lyapunov} with
$$p_{11}=0, \quad p_{12}=0, \quad p_{22}=1.$$
According to the law of conservation of energy, when there is no control input, the mechanical energy of a system remains constant. Therefore, the true time derivative of Lyapunov function of this system is zero. To verify this, we apply both our stability and instability verification algorithms to the system. They use the same data and Lipschitz constants which are obtained in a similar way to the cases of stability and instability verification. We visualize the time derivative of Lyapunov function estimated by stability and instability verification in Figure \ref{fig: pendulum critical v_dot}. The estimates obtained through stability verification are above zero in the entire state space while those given by instability verification are below zero. The estimates have opposite signs than required by $\eta$-testing stability criterion. Thus, our algorithm does not give conclusions about either the stability or instability of the system. Examining the estimated values further, we find that they are very close to zero compared with the results shown in Figure \ref{fig: pendulum stable v_dot} and \ref{fig: pendulum unstable v_dot}. This is consistent with the fact that the true time derivative of Lyapunov function is zero and is an important feature reflecting critical stability of datatic control systems.

\section{Conclusion}
This paper proposes a stability criterion for general datatic control systems called $\eta$-testing. Our criterion leverages the Lipschitz continuity assumption to extend information on known data points to unknown regions and restricts the unknown dynamics to the intersection of closed balls. We estimate the local Lipschitz constants by solving a QP and compute the worst-case time derivative of Lyapunov function inside the intersection by solving a QCLP. The relationship between the optimal values of QCLPs and zero gives conclusions about system stability. Experiments on both linear and nonlinear systems show that our algorithm successfully verifies the stability, instability, and critical stability of datatic control systems.

\bibliographystyle{IEEEtran}
\bibliography{ref}

\begin{thebibliography}{10}
\providecommand{\url}[1]{#1}
\csname url@samestyle\endcsname
\providecommand{\newblock}{\relax}
\providecommand{\bibinfo}[2]{#2}
\providecommand{\BIBentrySTDinterwordspacing}{\spaceskip=0pt\relax}
\providecommand{\BIBentryALTinterwordstretchfactor}{4}
\providecommand{\BIBentryALTinterwordspacing}{\spaceskip=\fontdimen2\font plus
\BIBentryALTinterwordstretchfactor\fontdimen3\font minus \fontdimen4\font\relax}
\providecommand{\BIBforeignlanguage}[2]{{%
\expandafter\ifx\csname l@#1\endcsname\relax
\typeout{** WARNING: IEEEtran.bst: No hyphenation pattern has been}%
\typeout{** loaded for the language `#1'. Using the pattern for}%
\typeout{** the default language instead.}%
\else
\language=\csname l@#1\endcsname
\fi
#2}}
\providecommand{\BIBdecl}{\relax}
\BIBdecl

\bibitem{aastrom2014control}
K.~J. {\AA}str{\"o}m and P.~R. Kumar, ``Control: A perspective.'' \emph{Automatica}, vol.~50, no.~1, pp. 3--43, 2014.

\bibitem{tsien1954engineering}
H.~Tsien, \emph{Engineering Cybernetics}.\hskip 1em plus 0.5em minus 0.4em\relax McGray-Hill, 1954.

\bibitem{wiener2019cybernetics}
N.~Wiener, \emph{Cybernetics or Control and Communication in the Animal and the Machine}.\hskip 1em plus 0.5em minus 0.4em\relax MIT press, 2019.

\bibitem{kalman1960contributions}
R.~E. Kalman, ``Contributions to the theory of optimal control,'' \emph{Bol. Soc. Mat. Mexicana}, vol.~5, no.~2, pp. 102--119, 1960.

\bibitem{zhou1998essentials}
K.~Zhou and J.~C. Doyle, \emph{Essentials of robust control}.\hskip 1em plus 0.5em minus 0.4em\relax Prentice hall Upper Saddle River, NJ, 1998, vol. 104.

\bibitem{mayne2014model}
D.~Q. Mayne, ``Model predictive control: Recent developments and future promise,'' \emph{Automatica}, vol.~50, no.~12, pp. 2967--2986, 2014.

\bibitem{li2004iterative}
W.~Li and E.~Todorov, ``Iterative linear quadratic regulator design for nonlinear biological movement systems,'' in \emph{First International Conference on Informatics in Control, Automation and Robotics}, vol.~2, 2004, pp. 222--229.

\bibitem{berberich2020data}
J.~Berberich, J.~K{\"o}hler, M.~A. M{\"u}ller, and F.~Allg{\"o}wer, ``Data-driven model predictive control with stability and robustness guarantees,'' \emph{IEEE Transactions on Automatic Control}, vol.~66, no.~4, pp. 1702--1717, 2020.

\bibitem{guan2021direct}
Y.~Guan, S.~E. Li, J.~Duan, J.~Li, Y.~Ren, Q.~Sun, and B.~Cheng, ``Direct and indirect reinforcement learning,'' \emph{International Journal of Intelligent Systems}, vol.~36, no.~8, pp. 4439--4467, 2021.

\bibitem{li2023reinforcement}
S.~E. Li, \emph{Reinforcement learning for sequential decision and optimal control}.\hskip 1em plus 0.5em minus 0.4em\relax Springer Singapore, 2023.

\bibitem{pradeep1990stability}
S.~Pradeep and S.~Shrivastava, ``Stability of dynamical systems-an overview,'' \emph{Journal of Guidance, Control, and Dynamics}, vol.~13, no.~3, pp. 385--393, 1990.

\bibitem{routh1877treatise}
E.~J. Routh, \emph{A treatise on the stability of a given state of motion, particularly steady motion: being the essay to which the Adams prize was adjudged in 1877, in the University of Cambridge}.\hskip 1em plus 0.5em minus 0.4em\relax Macmillan and Company, 1877.

\bibitem{hurwitz1895ueber}
A.~Hurwitz, ``Ueber die bedingungen, unter welchen eine gleichung nur wurzeln mit negativen reellen theilen besitzt,'' \emph{Mathematische Annalen}, vol.~46, no.~2, pp. 273--284, 1895.

\bibitem{nyquist1932regeneration}
H.~Nyquist, ``Regeneration theory,'' \emph{Bell System Technical Journal}, vol.~11, no.~1, pp. 126--147, 1932.

\bibitem{evans1948graphical}
W.~R. Evans, ``Graphical analysis of control systems,'' \emph{Transactions of the American Institute of Electrical Engineers}, vol.~67, no.~1, pp. 547--551, 1948.

\bibitem{lyapunov1992general}
A.~M. Lyapunov, \emph{General problem of the stability of motion}.\hskip 1em plus 0.5em minus 0.4em\relax CRC Press, 1992, vol.~55, no.~3.

\bibitem{bogoliubov1961asymptotic}
N.~N. Bogoliubov and Y.~A. Mitropolsky, \emph{Asymptotic methods in the theory of non-linear oscillations}.\hskip 1em plus 0.5em minus 0.4em\relax Gordon and Breach New York, 1961, vol.~1.

\bibitem{popov1961absolute}
V.~M. Popov, ``Absolute stability of nonlinear systems of automatic control,'' \emph{Avtomatika i Telemekhanika}, vol.~22, no.~8, pp. 961--979, 1961.

\bibitem{sandberg1965some}
I.~W. Sandberg, ``Some results on the theory of physical systems governed nonlinear functional equations,'' \emph{Bell System Technical Journal}, vol.~44, no.~5, pp. 871--898, 1965.

\bibitem{zames1966input}
G.~Zames, ``On the input-output stability of time-varying nonlinear feedback systems part one: Conditions derived using concepts of loop gain, conicity, and positivity,'' \emph{IEEE Transactions on Automatic Control}, vol.~11, no.~2, pp. 228--238, 1966.

\bibitem{willems2005note}
J.~C. Willems, P.~Rapisarda, I.~Markovsky, and B.~L. De~Moor, ``A note on persistency of excitation,'' \emph{Systems \& Control Letters}, vol.~54, no.~4, pp. 325--329, 2005.

\bibitem{de2019formulas}
C.~De~Persis and P.~Tesi, ``Formulas for data-driven control: Stabilization, optimality, and robustness,'' \emph{IEEE Transactions on Automatic Control}, vol.~65, no.~3, pp. 909--924, 2019.

\bibitem{van2020willems}
H.~J. van Waarde, C.~De~Persis, M.~K. Camlibel, and P.~Tesi, ``Willems’ fundamental lemma for state-space systems and its extension to multiple datasets,'' \emph{IEEE Control Systems Letters}, vol.~4, no.~3, pp. 602--607, 2020.

\bibitem{van2020data}
H.~J. Van~Waarde, J.~Eising, H.~L. Trentelman, and M.~K. Camlibel, ``Data informativity: a new perspective on data-driven analysis and control,'' \emph{IEEE Transactions on Automatic Control}, vol.~65, no.~11, pp. 4753--4768, 2020.

\bibitem{lavaei2022data}
A.~Lavaei, P.~M. Esfahani, and M.~Zamani, ``Data-driven stability verification of homogeneous nonlinear systems with unknown dynamics,'' in \emph{IEEE 61st Conference on Decision and Control}, 2022, pp. 7296--7301.

\bibitem{guo2021data}
M.~Guo, C.~De~Persis, and P.~Tesi, ``Data-driven stabilization of nonlinear polynomial systems with noisy data,'' \emph{IEEE Transactions on Automatic Control}, vol.~67, no.~8, pp. 4210--4217, 2021.

\bibitem{choi2021convex}
H.~Choi, U.~Vaidya, and Y.~Chen, ``A convex data-driven approach for nonlinear control synthesis,'' \emph{Mathematics}, vol.~9, no.~19, p. 2445, 2021.

\bibitem{markovsky2021behavioral}
I.~Markovsky and F.~D{\"o}rfler, ``Behavioral systems theory in data-driven analysis, signal processing, and control,'' \emph{Annual Reviews in Control}, vol.~52, pp. 42--64, 2021.

\bibitem{ye1989extension}
Y.~Ye and E.~Tse, ``An extension of karmarkar's projective algorithm for convex quadratic programming,'' \emph{Mathematical Programming}, vol.~44, pp. 157--179, 1989.

\bibitem{lobo1998applications}
M.~S. Lobo, L.~Vandenberghe, S.~Boyd, and H.~Lebret, ``Applications of second-order cone programming,'' \emph{Linear Algebra and Its Applications}, vol. 284, no. 1-3, pp. 193--228, 1998.

\end{thebibliography}

\end{document}